\documentclass{aa}

\usepackage{graphicx}
\usepackage{txfonts}
\usepackage{bm}
\usepackage{epstopdf}
\usepackage{multirow} 
\usepackage{rotating}
\usepackage{color}
\usepackage{xfrac}
\usepackage{natbib}
\usepackage{array} 
\usepackage{subcaption} 
 
 \definecolor{gris25}{gray}{0.75}
 
\newcolumntype{M}[1]{>{\centering\arraybackslash}m{#1}}
\newcolumntype{C}[1]{>{\centering\arraybackslash}m{#1}}
\newcolumntype{R}[1]{>{\raggedleft\arraybackslash}m{#1}}

\begin{document} 

   \title{Focal plane wavefront sensor achromatization :\\ The multireference self-coherent camera}
   \titlerunning{Focal plane wavefront sensor achromatization : The multireference self-coherent camera} 
   \authorrunning{J.Delorme}

   \author{J. R. Delorme\inst{1}
          \and
          R. Galicher\inst{1}
          \and          
          P. Baudoz \inst{1}
          \and
          G. Rousset\inst{1}
          \and          
          J. Mazoyer \inst{2}          
          \and
          O. Dupuis\inst{1}
          }

   \institute{LESIA, Observatoire de Paris, CNRS and University Denis Diderot Paris 7, 5 place Jules Janssen, 92195 Meudon, France
             \and
             Space Telescope Science Institute, 3700 San Martin Drive, 21218 Baltimore MD, USA
             \\ \email{jacques-robert.delorme@obspm.fr}
             }

   \date{Received  28 October 2015 / Accepted 5 January 2016}
 
  \abstract
   {High contrast imaging and spectroscopy provide unique constraints for exoplanet formation models as well as for planetary atmosphere models. But this can be challenging because of the planet-to-star small angular separation ($<1$ arcsec) and high flux ratio ($>10^{5}$). Recently, optimized instruments like VLT/SPHERE and Gemini/GPI were installed on 8m-class telescopes. These will probe young gazeous exoplanets at large separations ($\gtrsim$ 1au) but, because of uncalibrated phase and amplitude aberrations that induce speckles in the coronagraphic images, they are not able to detect older and fainter planets.}
   {There are always aberrations that are slowly evolving in time. They create quasi-static speckles that cannot be calibrated a posteriori with sufficient accuracy. An active correction of these speckles is thus needed to reach very high contrast levels ($>10^{6} - 10^{7}$). This requires a focal plane wavefront sensor. Our team proposed a self coherent camera, the performance of which was demonstrated in the laboratory. As for all focal plane wavefront sensors, these are sensitive to chromatism and we propose an upgrade that mitigates the chromatism effects.}
   {First, we recall the principle of the self-coherent camera and we explain its limitations in polychromatic light. Then, we present and numerically study two upgrades to mitigate chromatism effects: the optical path difference method and the multireference self-coherent camera. Finally, we present laboratory tests of the latter solution. }
   {We demonstrate in the laboratory that the multi-reference self-coherent camera can be used as a focal plane wavefront sensor in polychromatic light using an 80 nm bandwidth at 640 nm (bandwidth of 12.5\%). We reach a performance that is close to the chromatic limitations of our bench: 1$\sigma$ contrast of $4.5\: 10^{-8}$ between 5 and 17 $\lambda_{0}/D$.}
   {The performance of the MRSCC is promising for future high-contrast imaging instruments that aim to actively minimize the speckle intensity so as to detect and spectrally characterize faint old or light gaseous planets.}
   \maketitle

\section{Introduction}

During the last few years, the imaging of circumstellar disks and exoplanets has become one of the priorities to be able to constrain models of planetary system formation and of planetary atmospheres. However, direct imaging of faint sources around bright objects is very challenging. For example, in the visible and near-infrared light, the contrast ranges from $10^{3}$ for bright debris disks down to $10^{10}$ for Earth-like planets \citep{Seager2010}. The study of this type of object therefore requires dedicated techniques, such as coronagraphs, that attenuate the light from the host star \citep[][]{Marois2010, Lagrange2009}. 

A lot of coronagraphs associate a focal plane mask and a pupil diaphragm called a Lyot stop to filter the stellar light and minimize the star intensity in the science image \citep{Rouan2000,Soummer2003,Mawet2005,Murakami2008}. However, as a result of aberrations, part of the stellar light goes through the Lyot stop and produces speckles in the science image. These speckles are usually brighter than the planet signal that we are looking for and it is necessary to reduce their intensity. 

With ground-based telescopes, most of the speckles are due to atmospheric turbulence. These dynamic wavefront errors are estimated and corrected by conventional adaptive optics (AO) systems that measure the wavefront error using a wavefront sensor in a dedicated optical channel. Because of beam splitting, quasi-static, non-common path aberrations (NCPA) are generated by the instrument optics between the wavefront-sensing channel and the science-image channel, which limits the accuracy of the AO correction \citep{Hartung2003}. Like P1640 \citep{Hinkley2010} and SCExAO \citep{Jovanovic2014},  VLT/SPHERE \citep{Beuzit2008} and Gemini/GPI \citep{Macintosh2008} are two instruments designed to detect young Jupiter-like planets. They minimize the quasi-static NCPA in open loop before the observations but cannot control them during the observations. However, as NCPAs evolve during the observations, the calibration degrades with time and the contrast performance is limited by quasi-static speckles to $10^{-4} - 10^{-5} $ at 0.5'' in raw images. Space-based telescopes are free of atmospheric turbulence but the variations of temperature and gravity also create quasi-static aberrations. Dedicated real-time methods are required to estimate and compensate for quasi-static aberrations in a closed loop. First, these can be implemented into current ground-based high-contrast imaging instruments to enhance the contrast in the science images. Secondly, we could integrate these techniques in the design of high-contrast instruments for future ground-based extremely large telescopes (ELTs) and space-based telescopes, e.g. WFIRST-AFTA \cite{Spergel2015} or HDST, see \cite{Dalcanton2015}. 

Dedicated strategies for observations \citep{Marois2004,Marois2006} have been implemented to overcome the quasi-static speckle limitation. They use the correlation between speckles in time or in wavelength. Their performance were demonstrated on-sky, enabling exoplanets to be detected by direct imaging \citep{Marois2008,Marois2010,Lagrange2009}. However, as the aberrations and the speckle pattern slowly evolve in time, a posteriori calibrations cannot calibrate all the speckles and an active control of the speckle field is requiered to achieve contrast that is better than $10^{-5}$ in the raw data.

This type of control can be done using at least one deformable mirror (DM) upstream from the coronagraph. Since DMs have a finite number of actuators, they can only correct speckles in a bounded area in the focal plane image \citep[Section \ref{SubSec_DH} and ][]{Malbet1995}. To correct the speckles without being limited by NCPA,  we also need to estimate the aberrations (i.e., the complex electric field of the speckles) directly from the science image using a focal plane wavefront sensor (FPWFS).  

Several FPWFSs have been proposed. Some of them are based on temporal modulations of speckles like the electrical field conjugation method \citep{Giveon2006,Giveon2007b,Borde2006} or phase diversity method \citep{Sauvage2012,Paul2013}. There are also FPWFSs that are based on spatial modulations of the speckles such as the Asymetric Pupil Fourier wavefront sensor \citep{Martinache2013} or the self-coherent camera \citep[SCC,][]{Baudoz2006,Galicher2008}. The SCC creates Fizeau interference to spatially modulate the speckle intensity in the coronagraphic images. Thus, it can retrieve the complex electric field from one coronagraphic image (i.e., the science image) without NCPA. We discuss the SCC principle and performance \citep{Mazoyer2013b} in Section \ref{Sec_SCC}. As with all FPWFSs, the SCC is limited by the chromatism of the images. In this paper, we propose two methods to make it work in polychromatic light. The first method is an upgrade of the SCC in which the optical path difference (OPD) between the two beams of the Fizeau interferometer is not null. The second solution, called the multireference self-coherent camera (MRSCC), is another upgrade of the SCC, which uses more than two interfering beams. We present both methods and discuss their performance and limitations in Sections \ref{Sec_piston} and \ref{Sec_MRSCC}. In Section \ref{Sec_THD_description}, we present the très haute dynamique (THD) laboratory bench and the expected performance of the MRSCC. Finally, in Section \ref{Sec_THD}, we demonstrate in the laboratory that the MRSCC can control, in closed loop, a DM that minimizes the speckle intensity in wide spectral band images (12.5\% bandpass).

\section{Dark hole and estimation of the contrast}
\label{SubSec_DH}
A DM has a limited number of actuators and only modify speckles in a bounded area of the image that we call  the influence area. Assuming an \textit{N} by \textit{N} actuator DM in pupil plane, the size of the influence area is a $N\lambda_{0}/D$ by $N\lambda_{0}/D$ square centered on the optical axis, where $\lambda_{0}$ is the central wavelength and D the diameter of the pupil. The part of the influence area where we try to minimize the speckle intensity is called dark hole \citep[DH,][]{Malbet1995}. If only one DM is used and set in a pupil plane, we can choose to control only the phase aberrations to reduce the speckle intensity in the full influence area, or both phase and amplitude aberrations by limiting the DH to half of the influence area \citep{Borde2006}. Because the aberrations are never solely phase aberrations in the experimental set up, we use the second option in this paper.

Once the speckle minimization have been performed, we estimate its performance in a contrast computation area. We define this area within the DH removing $1.5\lambda_{0}/D$ on each side. Indeed, since we do not use a pupil apodizer, the diffraction pattern of the uncorrected bright speckles that are located just outside of the DH spreads light inside the DH. 

\begin{figure}
        \centering
        \begin{subfigure}[b]{0.48\textwidth}        
        \includegraphics[width = \textwidth]{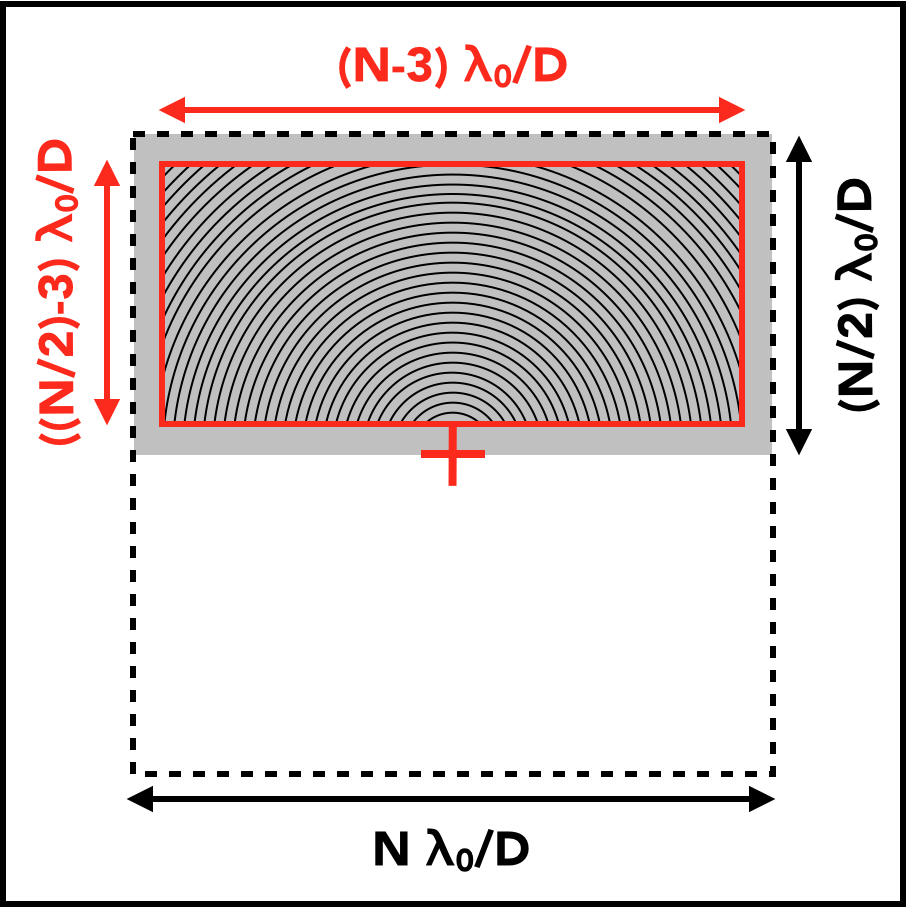}         
        \end{subfigure}
        \caption{Influence area of the DM (dashed line), DH (gray area) and computation area (red solid line). The annuli of $\frac{\lambda_{0}}{2D}$ width in the computation area are used to compute the contrast curves. The red cross represents the optical axis.}
        \label{DH}%
\end{figure}

Figure \ref{DH} is a scheme of the main sizes of the three areas that have been introduced previously. The influence area is delineated by a dashed line, with the DH represented as a gray area and the computation area delineated by a red solid line. The computation area covers a range of  separations from $1.5\lambda_{0}/D$ to $\left((N-3)/\sqrt{2}\right)\lambda_{0}/D$.

To estimate by how much the speckle intensity is reduced after correction, we use two criteria: the contrast curve and the cumulative function of the intensity, both of which are computed in the computation area. The contrast curves are the azimuthal standard deviation of the image intensity, computed in annuli of $\lambda_{0}/2D$ width (annuli in Fig. \ref{DH}) centered on the optical axis (red cross). It gives the 1$\sigma$ detection limit as a function of the angular separation. To obtain cumulative functions, we compute the ratio between the number of pixels with an intensity lower than a given value and the total number of pixels within the computation area. These two criteria are complementary. The cumulative function gives information about the statistics of the residual intensity (median and dispersion) while the contrast curve shows how the standard deviation evolves with the angular separation to the star.

\section{The self-coherent camera}
\label{Sec_SCC}

In this section, we recall the principles of the SCC and explain the origin of its sensitivity to chromatism.

\subsection{Principle of the SCC in monochromatic light}
\label{Sec_SCC_sub_mono}

\begin{figure}
        \centering
       \begin{subfigure}[b]{0.36\textwidth}
        \includegraphics[width = \textwidth]{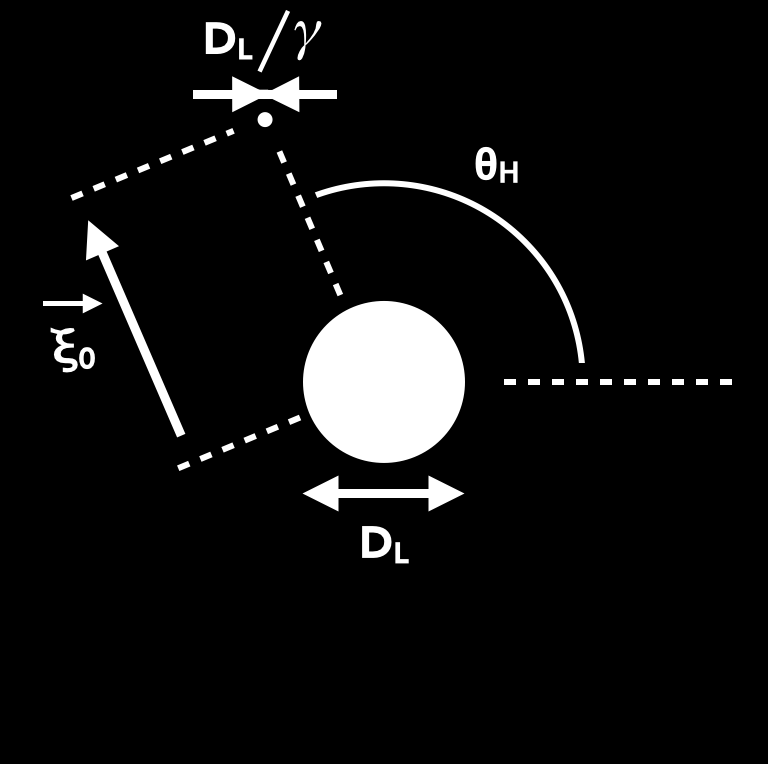}                
        \end{subfigure}

        \begin{subfigure}[b]{0.36\textwidth}
        \includegraphics[width = \textwidth]{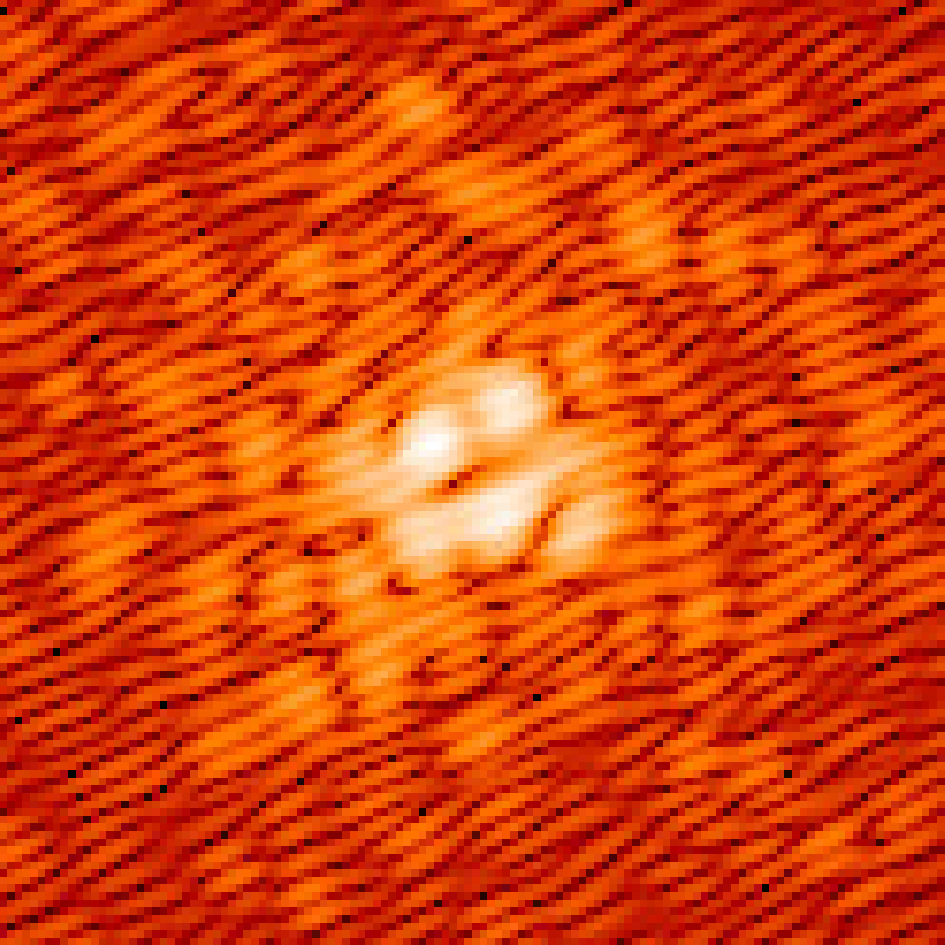}        
        \end{subfigure}
        
        \begin{subfigure}[b]{0.36\textwidth}
        \includegraphics[width = \textwidth]{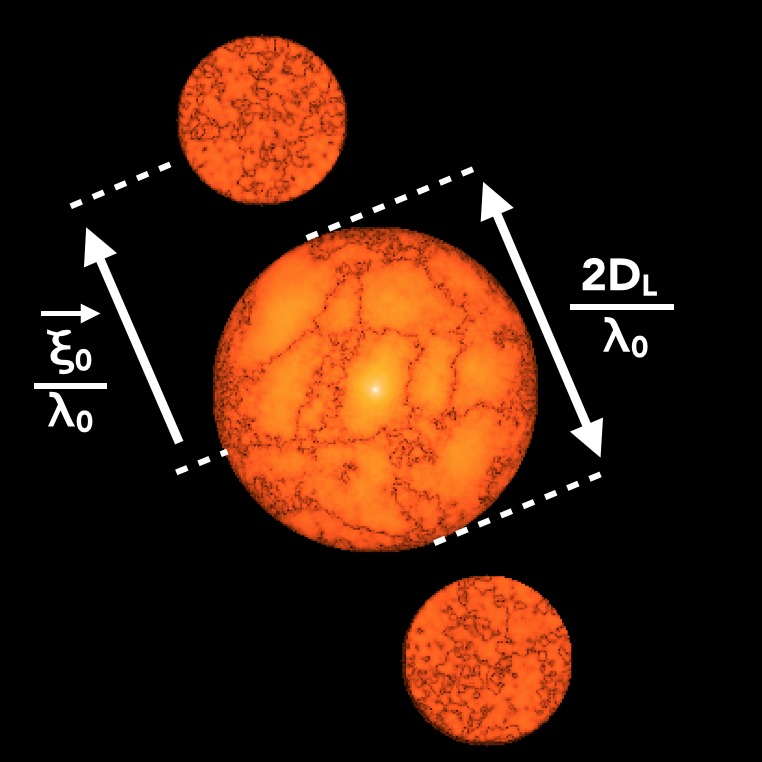}
        \end{subfigure}
        \caption{Top: SCC Lyot stop. Middle: Numerically simulated SCC image showing spatially modulated stellar speckles. Field of view: 20$\lambda_{0}/D$ by 20$\lambda_{0}/D$. Bottom: Fourier transformation of the SCC image.}
        \label{Fig_A}
\end{figure}

The SCC can be used as an FPWFS downstream from a coronagraph for high contrast imaging  \citep{Galicher2008,Mazoyer2014}. It uses a small reference hole that is added to the Lyot stop plane (Fig. \ref{Fig_A}, top). We call the separation between the classical Lyot stop and the reference hole $\overrightarrow{\xi_{0}}$, $D_{L}$ the diameter of the classical Lyot stop, and $\gamma$ the ratio between $D_{L}$ and the diameter of the reference hole. We consider $\gamma \gg 1$ \citep{Galicher2010}. The angle $\theta_{H}$ between the reference hole and the horizontal axis of the Lyot stop plane does not impact on the performance of the instrument. We use  an angle of +115$^{\circ}$ in this paper, which is the value for our laboratory experiments (Section \ref{Sec_THD_description}).

Owing to aberrations, part of the stellar light is not filtered by the coronagraph and goes through the classical Lyot stop, which induces speckles in the image. A small part of the stellar light that is diffracted by the coronagraph is selected by the SCC reference hole.
The two coherent beams are recombined in the image plane, which form Fizeau fringes that spatially modulate the speckles. As for a planet, its image is off the coronagraph mask and its light only goes through the classical Lyot stop. As it is not coherent with the stellar light of the reference hole, the companion image is not spatially modulated by fringes. Figure \ref{Fig_A} (middle) is an SCC image with no planet. In this image, we artificially increased the flux from the reference to underline the fringes.

We can write the intensity $I$ in an SCC image as follows:
\begin{eqnarray}
I(\overrightarrow{\alpha}) &=& \left\vert A_{S}\left(\overrightarrow{\alpha}\right)\right\vert^{2} + \left\vert A_{C}\left(\overrightarrow{\alpha}\right)\right\vert^{2} + \left\vert A_{R}\left(\overrightarrow{\alpha}\right)\right\vert^{2}  \nonumber \\
&+& 2 Re \left[ A_{S}\left(\overrightarrow{\alpha}\right)A_{R}^{*}\left(\overrightarrow{\alpha}\right) \exp\left( \frac{2 i \pi \overrightarrow{\alpha} . \overrightarrow{\xi_{0}}}{\lambda_{0}}\right)\right],
\label{Im_SCC_mono}
\end{eqnarray}
where $\overrightarrow{\alpha}$ is the focal plane coordinate. $A_{S}$ is the complex electric field of the speckles we want to estimate and minimize, $A_{C}$ the complex electric field of the companion and $A_{R}$ the complex electric field associated with the SCC reference hole and $A_{R}^{*}$ the conjugate of $A_{R}$. These complex electric fields in the focal plane are wavelength-dependent and we account for these dependences in all numerical simulations. In the second member of Equation \ref{Im_SCC_mono}, the first term is the speckle pattern (i.e., coronagraphic residue) that we want to minimize, the second term is the companion image that we want to detect, the third term is the intensity associated with the SCC reference hole and, finally, the last term corresponds to the spatial modulation of the speckles.

Once the image is recorded, we apply a numerical Fourier transform (Fig. \ref{Fig_A}, bottom) to differentiate the fringed speckles from the companion image and we obtain: 
\begin{eqnarray}
\mathcal{F}^{-1} \left[I\right](\overrightarrow{u}) &=&  \mathcal{F}^{-1} \left[ \vert A_{S} \vert ^{2} + \vert A_{C}\vert ^{2} + \vert A_{R}\vert ^{2} \right] * \delta \left( \overrightarrow{u} \right)\nonumber \\
&+&  \mathcal{F}^{-1} \left[A_{S}^{*}A_{R}\right] * \delta \left( \overrightarrow{u} +  \frac{\overrightarrow{\xi_{0}}}{\lambda_{0}} \right)  \nonumber \\   
&+& \mathcal{F}^{-1} \left[A_{S}A_{R}^{*}\right] * \delta \left( \overrightarrow{u} -  \frac{\overrightarrow{\xi_{0}}}{\lambda_{0}} \right) ,
\label{FT_SCC_mono}
\end{eqnarray}
where $\overrightarrow{u}$ is the spatial frequency plane coordinate, $\mathcal{F}$ the Fourier transform operator and $\mathcal{F}^{-1}$ its inverse. The first term, which is the central peak of the Fourier transform, is the sum of the autocorrelations of the electric field in the classical Lyot stop and in the reference hole. The radius of the central peak is  $D_{L}/\lambda_{0}$ because $\gamma \gg 1$. The other two terms, which correspond to the two lateral peaks of the Fourier transform, are the correlation between the stellar electric fields in the classical Lyot stop and in the reference hole. The two peaks are conjugated and contain the same information and the same contribution of noise. The radius of these peaks is $D_{L}\left(1+1/\gamma\right)/(2\lambda_{0})$. Thus, the three peaks do not overlap if the separation $\vert\vert \overrightarrow{\xi_{0}} \vert\vert$ between the classical Lyot stop and the SCC reference hole is large enough  \citep{Galicher2010}:
\begin{eqnarray}
\vert\vert \overrightarrow{\xi_{0}} \vert\vert &>& \frac{D_{L}}{2} \left( 3 + \frac{1}{\gamma}\right) .
\label{Eq_distance_mono}
\end{eqnarray}

To estimate the complex electric field $A_{S}$, we select one of the lateral peaks and we recenter it. The Fourier Transform of this correlation peak, called $I_{-}$, is
\begin{eqnarray}
I_{-} & = & A_{S}A_{R}^{*} .
\label{Centered_correlation_peak}
\end{eqnarray}

This contains the information on the complex electric field $A_{S}$ of the stellar speckles multiplied by the reference $A^{*}_{R}$. Because the reference beam is created in the Lyot plane of the instrument and goes through the same optics as the beam of the classical Lyot stop, $A^{*}_{R}$ is very stable with respect to $A_{S}$. Moreover, the point spread function (PSF) $\vert A_{R}\vert ^{2}$ has a full width at half maximum (FWHM) of $\gamma \: \lambda_{0}/D $. So, if the reference pupil is small enough, ($\gamma \gg 1$) $\vert A_{R}\vert ^{2}$ does not go to zero inside the influence area of the DM. With these hypotheses, if we minimize $I_{-}$, as done in this paper, we also minimize $A_{S}$ and the speckle intensity \citep{MazoyerPHD}.

To control the DM on our bench, we reform the real and imaginary parts of $I_{-}$ as vectors and we concatenate them in one unique vector. To convert $I_{-}$ in a command to the DM we make use of a linear approach that is based on a control matrix \citep{Boyer1990,Mazoyer2013a}. To build the control matrix, we define the set of sine and cosine functions that the DM can produce. The number of functions equals the number of degrees of freedom available with the DM. We apply each of these functions successively to the DM and record the corresponding SCC image and compute the corresponding vector $I_{-}$. With all the $I_{-}$, we build an interaction matrix. Finally, we invert this using a singular value decomposition to obtain the control matrix that can be used in a closed loop.

During the correction, we register one SCC image at each iteration. To only correct speckles in the DH (and not in the entire influence area), we multiply the SCC image by a numerical mask (Butterworth type) reproducing the DH. 
The resulting image is processed to extract $I_{-}$. The numerical mask allows us to be able to saturate part of the image outside the DH with limited impacts on the estimated $I_{-}$. The multiplication of $I_{-}$ by the control matrix leads to the DM command increment that is multiplied by a gain and added to the previous iteration command. After a few iterations the correction has been done and is stable.

Usually, the active correction is not limited by the SCC and the speckles inside the DH are still modulated by fringes. We can then apply the a posteriori SCC data processing to enhance the contrast \citep{Baudoz2013}. In this paper we do not use the a posteriori calibration and we focus on the performance of the active correction. Consequently, after the active correction is done, we can close the reference aperture to record the coronagraphic image (with no stellar light from the reference hole). We normalize the SCC and coronagraphic images by the maximum intensity of the non-coronagraphic PSF that was recorded with the stellar source off the coronagraph axis. All the images shown in this paper are normalized in flux. 

\subsection{Simulated performance in monochromatic light}
\label{Sec_SCC_sub_perf_mono}

Below we derive a typical performance for the SCC in monochromatic light ($\lambda_{0} = 640\:\textrm{nm}$) from numerical simulations. This result is used when studying the performance in polychromatic light (Section \ref{subsec_SCCpolyperf}). 

We assume a perfect achromatic coronagraph \citep{Cavarroc2006} without any apodization. To simulate the SCC, we numerically add a reference hole, assuming $\gamma = 25$ and $\xi_{0} = 1.8D$, which obey Equation \ref{Eq_distance_mono}. The flux inside of the reference beam is similar to the flux measured in our laboratory, which is $4.8\:10^{-8}$  times the energy in the pupil upstream from the coronagraph. The simulated phase aberrations have a power spectral density (PSD), which varyies as $f^{-2.3}$ (with f the spatial frequencies) and a standard deviation of 10 nm RMS inside the pupil, since it is the case in our laboratory experiment. 
We also assume amplitude aberrations of 5\% RMS with a PSD, which varies as $f^{-1}$. Levels and PSD chosen for phase and amplitude are similar to those we measured in our laboratory. We assume a 32 by 32 actuator DM and we correct phase and amplitude aberrations in a DH of 16 by 32 $\lambda_{0}/D$. The simulated detector is similar to the one that we used in our laboratory. This has a readout noise (RON) of 3.2 e$^{-}$ per pixel  and a full well capacity of 60,000 e$^{-}$ per pixel.

\begin{table}
\centering
\begin{tabular}{p{3.2cm} p{4.9cm}}
\hline
\hline
Light source &  $\lambda_{0} = 640\:\textrm{nm}$ \\
\hline
Phase aberrations &  PSD: $f^{-2.3}$, 10 nm RMS \\
\hline
Amplitude aberrations & PSD: $f^{-1}$, 5\% RMS \\
\hline
Deformable mirror  &  32 by 32 actuators \\
\hline
Coronagraph  &  Perfect coronagraph\\
\hline
Lyot stop  & $\gamma$ = 25, $\xi_{0}$ = 1.8 $D_{L}$, $\theta_{H} = +115^{\circ}$ \\
\hline
Readout noise & 3.2 e$^{-}$/pixel\\
\hline
Saturation level & 60,000 e$^{-}$/pixel  \\
\hline
Normalized intensity & $2\:10^{-8} \longleftrightarrow 17 $ e$^{-}$/pixel\\
\hline
DM influence area & 32 by 32 $\lambda_{0}/D$ \\
\hline
Dark hole & 16 by 32 $\lambda_{0}/D$ \\
\hline
Computation area & 13 by 29 $\lambda_{0}/D$ \\
\hline
Sampling & 6.25 pixels per $\lambda_{0}/D$\\
\hline
\hline
\end{tabular}
\caption{Set of parameters used to simulate the image of Fig. \ref{fig:mono_HDH:Im_coro}.} 
\label{Tab_B}
\end{table}

In this paper, we reproduce the readout noise of the detector and the photon noise. To set the flux level, we use our experimental conditions  (see Section \ref{Sec_THD}), with 17 e$^{-}$ for a normalized intensity of $2.\:10^{-8}$. We use this conversion factor in the numerical simulations. We also reproduce the saturation of the detector. All the corrected images presented in this paper are coronagraphic images.

\begin{figure}
        \centering
        \begin{subfigure}[b]{0.36\textwidth}
        \includegraphics[width = \textwidth]{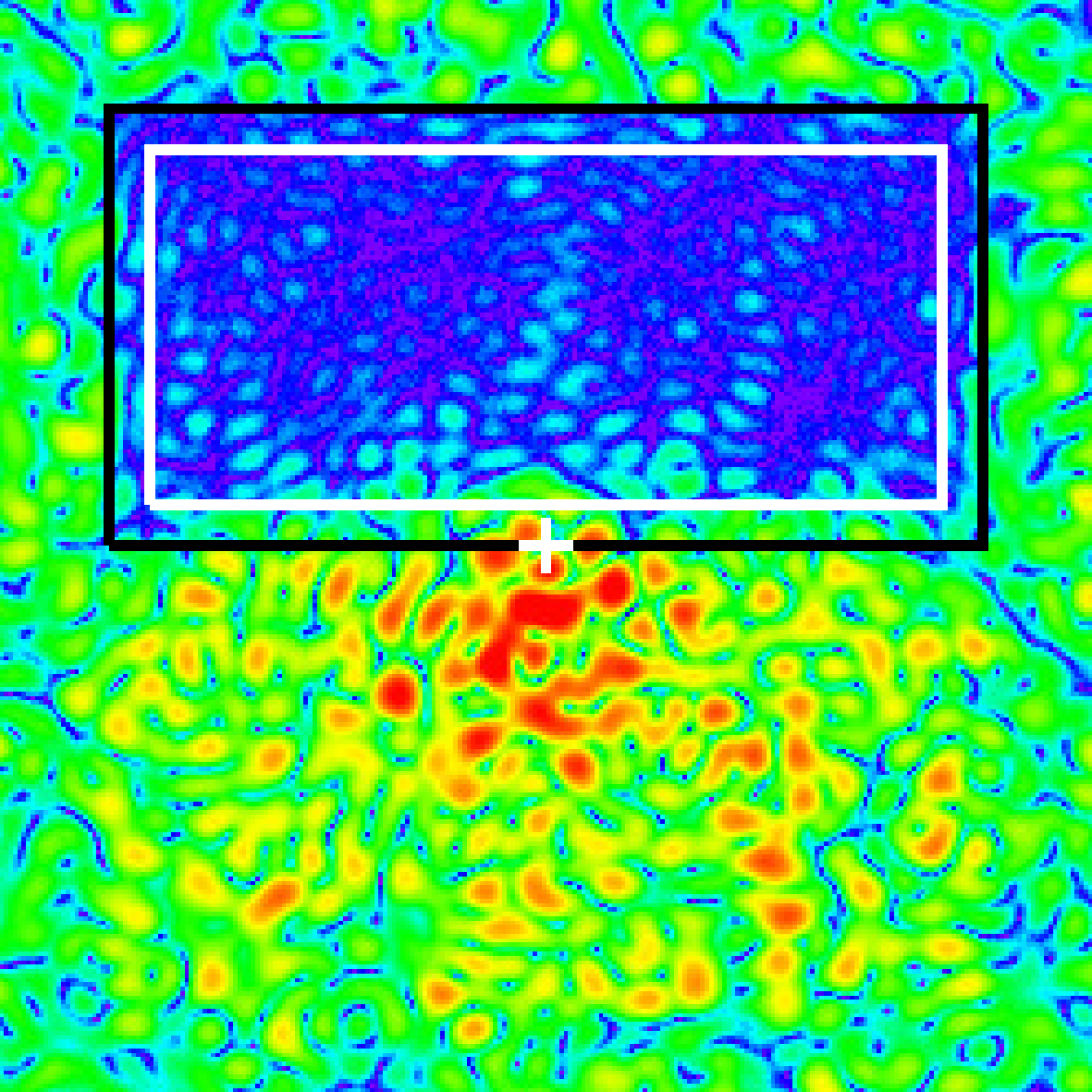}                        
        \end{subfigure}
        \begin{subfigure}[b]{0.098\textwidth} 
        \includegraphics[width = \textwidth]{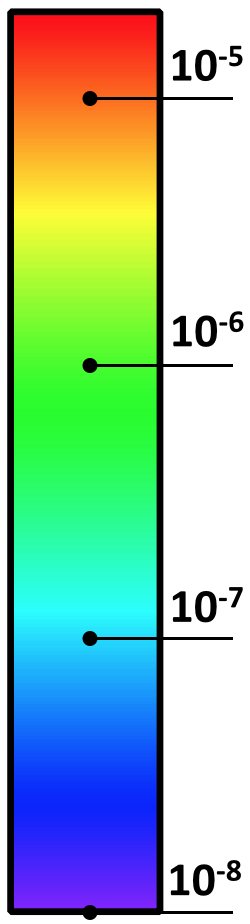}
        \end{subfigure}        
        \caption{Numerically simulated coronagraphic image obtained after an SCC correction in monochromatic light. Simulation parameters are given in Table \ref{Tab_B}. Black line: DH (16 $\lambda_{0}/D$ by 32 $\lambda_{0}/D$). White line: computation area (13 $\lambda_{0}/D$ by 29 $\lambda_{0}/D$). White cross: optical axis. Field of view: 40 $\lambda_{0}/D$ by 40 $\lambda_{0}/D$. The color bar associated with this image is the same for all the coronagraphic images in the paper.}
        \label{fig:mono_HDH:Im_coro}
\end{figure}

Figure \ref{fig:mono_HDH:Im_coro} presents the coronagraphic image obtained after correction. In monochromatic light, the intensity of all the speckles located inside the computation area (white line) is efficiently reduced. The perfect coronagraph that we simulated does not include any apodization. As a consequence, the bright speckles close to optical axis (white cross) spread light at small angular separations in the DH.

\begin{figure}
        \centering
        \begin{subfigure}[b]{0.36\textwidth}                
        \includegraphics[trim = 11mm 6mm -10mm 8mm, clip, width = 1.1 \textwidth]{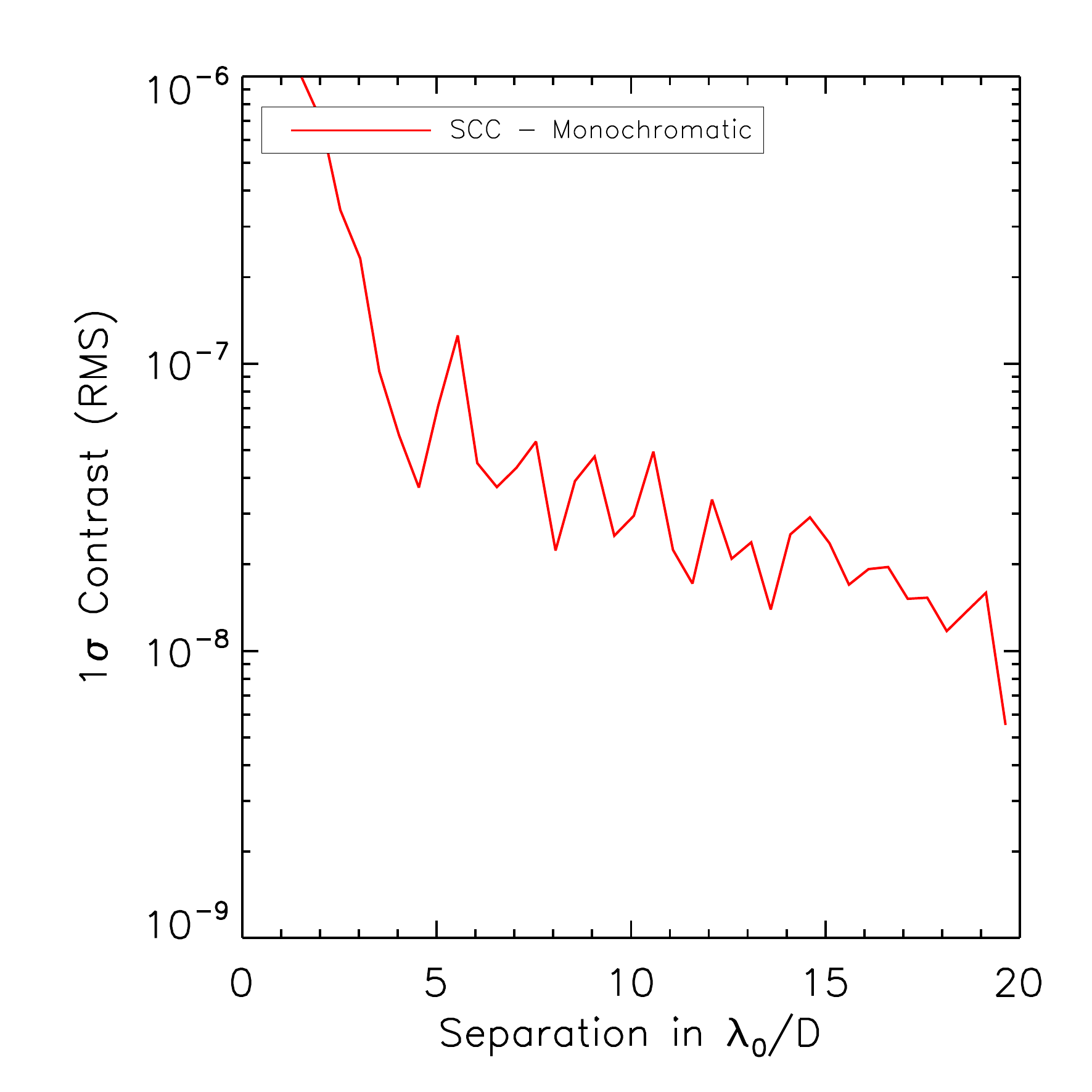}                              
        \end{subfigure}
        
        \begin{subfigure}[b]{0.36\textwidth}                
        \includegraphics[trim = 11mm 6mm -10mm 8mm, clip, width = 1.1 \textwidth]{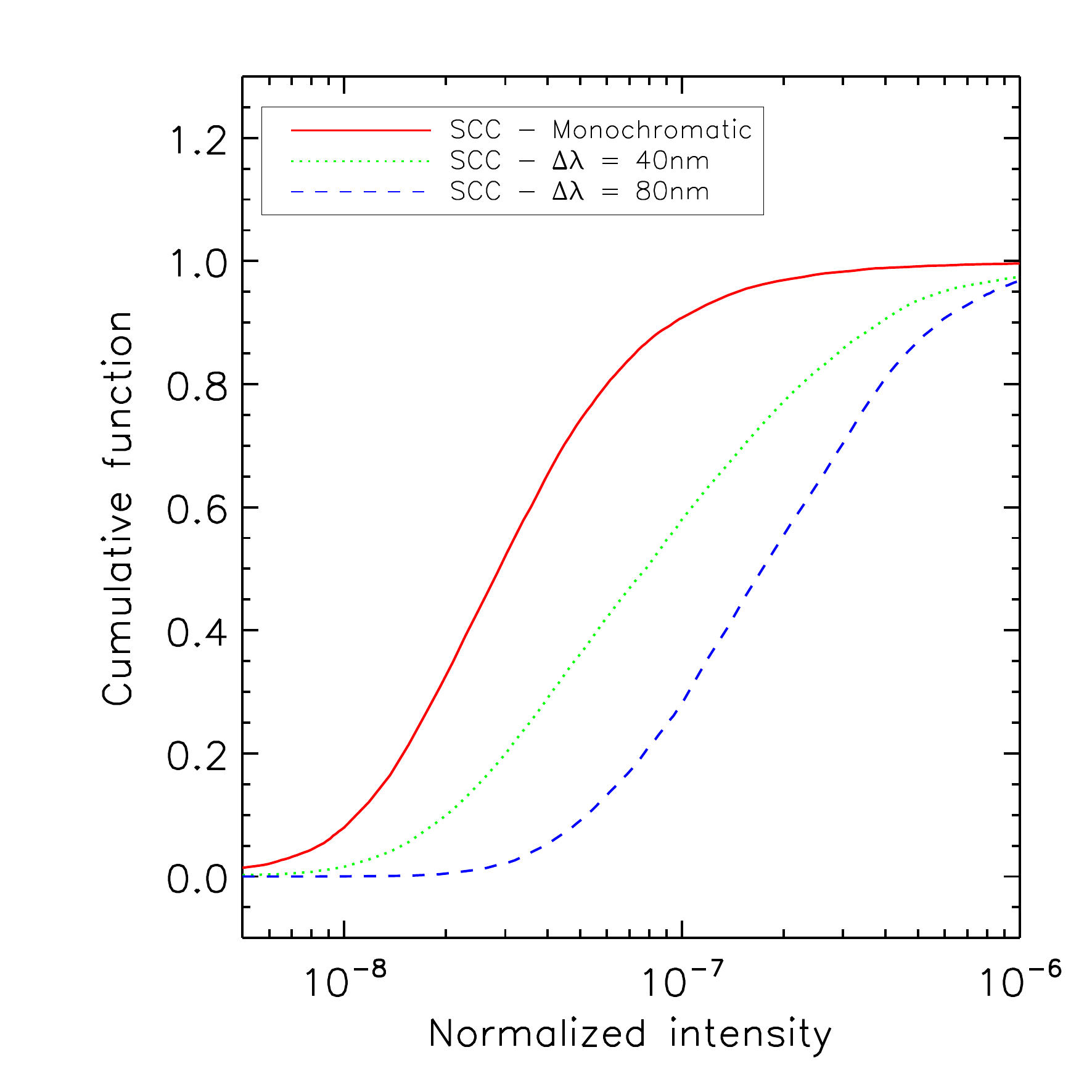}     
        \end{subfigure}        
        \caption{Contrast curve (top) computed from the coronagraphic image that was obtained with the SCC in monochromatic light (Fig. \ref{fig:mono_HDH:Im_coro}) and spatial cumulative function (bottom) associated with images obtained with the SCC in monochromatic light (red solid line) and in polychromatic light, assuming $\Delta\lambda = 40\:\textrm{nm}$ (green dotted line) and $80\:\textrm{nm}$ (blue dashed line).}  
        \label{Fig_results_scc_mono_PC}
\end{figure}
Figure \ref{Fig_results_scc_mono_PC} presents the contrast curve (top) and the cumulative function (bottom) associated with the image of Fig. \ref{fig:mono_HDH:Im_coro}. The average of the 1$\sigma$ contrast is $2.9\:10^{-8}$ between $5\lambda_{0}/D$ and $20\lambda_{0}/D$ and between $6\lambda_{0}/D$ and $17\lambda_{0}/D$. From the cumulative function the normalized intensity is better than $2.8\:10^{-8}$ in 50\% of the computation area (i.e., the median-normalized intensity). Moreover, the cumulative curve shows that the dispersion of the speckle intensity is small as observed in Fig. \ref{fig:mono_HDH:Im_coro} where the contrast level inside the DH is almost uniform. The achieved contrast level depends on the level of phase and amplitude aberrations assumed in the simulation. 

\cite{Mazoyer2014} demonstrated that the limitation of the correction obtained in monochromatic light in the laboratory was that the amplitude aberrations cannot be corrected with only one deformable mirror. In other words, the SCC electric field estimation is accurate but the correction with a finite number of actuators is limited. The results presented here assumed a clear circular aperture. For a pupil with central obscuration, spiders, or segmentation, as long as the coronagraph rejects part of the stellar light in the reference hole, the SCC should then provide an accurate estimation of the electric field that needs to be corrected. However, discountinuities in the pupil produce effects in the focal plane that could be addressed using dedicated coronagraphs \citep[e.g.,][]{NDiaye2014} or special correction techniques with several DMs in cascade, see \cite{Pueyo2013}.

\subsection{SCC in polychromatic light}
\label{subsec_SCCpolyformalism}

To spectrally characterize the very faint neighborhood of bright stars like exoplanets and increase the signal to noise ratio of the detections, one solution is to make observations with a large bandwidth (5-20\%). As a consequence, an efficient coronagraphic instrument has to work with large bandwidths as well as the associated focal plane wavefront sensor. In this section, we assume a perfectly achromatic coronagraph to study the impact of a finite bandwidth on the SCC wavefront estimation.

\begin{figure}
        \centering
        \begin{subfigure}[b]{0.24\textwidth}
                \includegraphics[width = \textwidth]{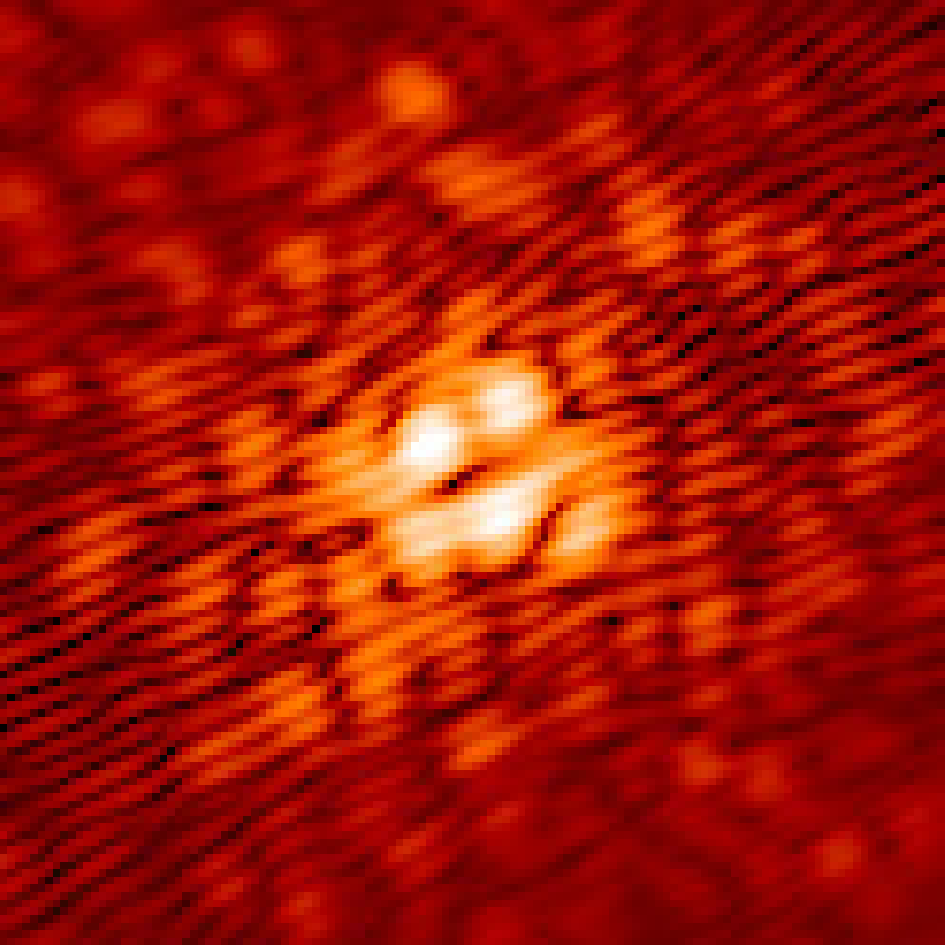}
        \end{subfigure}
        \hspace{-0.1cm}
        \begin{subfigure}[b]{0.24\textwidth}        
                \includegraphics[width = \textwidth]{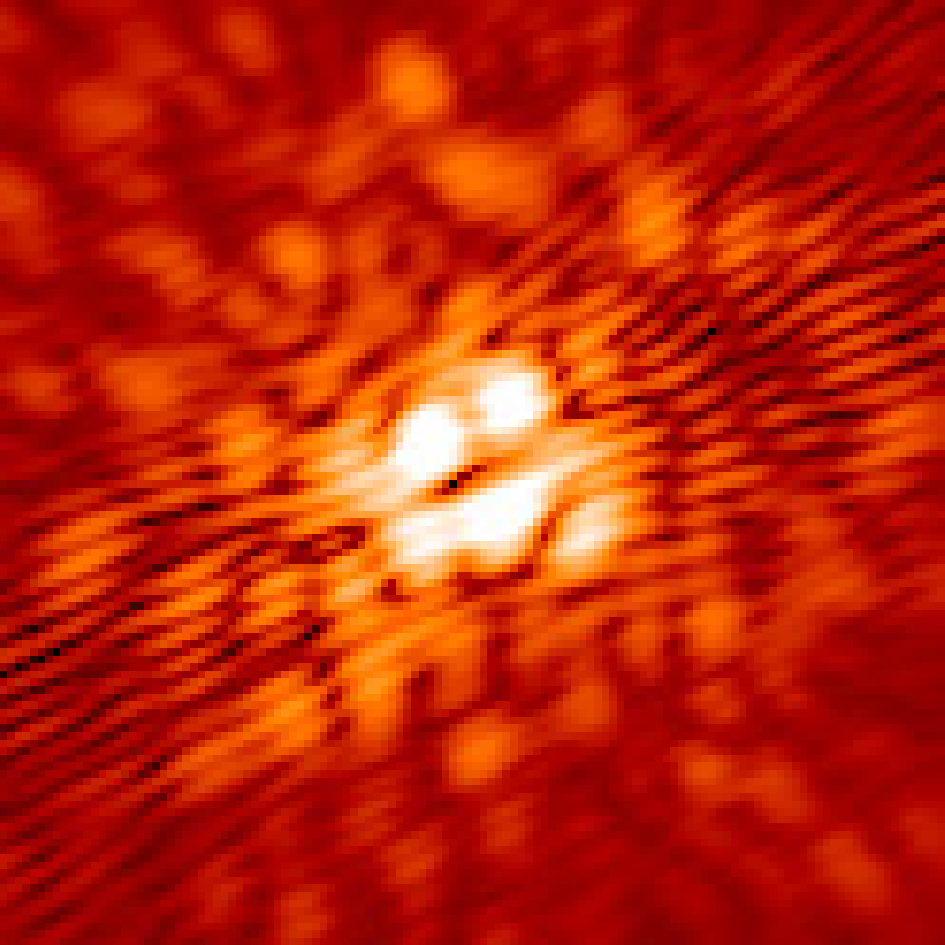}
        \end{subfigure}
        \caption{Numerically simulated SCC images obtained before correction for $\Delta\lambda=40\:\textrm{nm}$ (left) and $\Delta\lambda=80\:\textrm{nm}$ (right) at $\lambda_{0} = 640\:\textrm{nm}$. Field of view: 20 by 20 $\lambda_{0}/D$.}
        \label{Fig_SCC_poly_before_correction}
\end{figure}

Figure \ref{Fig_SCC_poly_before_correction} presents SCC images obtained before correction, using the same parameters as for the monochromatic case (Table \ref{Tab_B}) but with larger bandwidths:  $\Delta\lambda = 40\:\textrm{nm}$ (left) and $\Delta\lambda = 80\:\textrm{nm}$ (right). As for Fig. \ref{Fig_A} (middle), to highlight the fringes, we numerically increased the flux into the reference beams to obtain these two SCC images. We observe that the fringes are only located around the central fringe with a width that is proportional to the coherence length of the light $(\lambda_{0}^{2}/\Delta\lambda)$. Indeed, in polychromatic light, the intensity recorded on the detector in the focal plane is 

\begin{eqnarray}
I_{\Delta\lambda}(\overrightarrow{\alpha}) &=&  \int_{\Delta\lambda}  \left[\vert A_{S} \vert ^{2} + \vert A_{C}\vert ^{2} + \vert A_{R}\vert ^{2} \right]d\lambda  \nonumber \\
&+& \int_{\Delta\lambda} 2Re \left[ A_{S}A_{R}^{*}\exp\left( \frac{2 i \pi \overrightarrow{\alpha} . \overrightarrow{\xi_{0}}}{\lambda}\right)\right]d\lambda , 
\label{Im_SCC_poly}
\end{eqnarray}
where $\int_{\Delta\lambda}$ represents the integration over the range $\lambda_{0} \pm \Delta\lambda/2$. As in Equation \ref{Im_SCC_mono}, the last term corresponds to the spatial modulation of the speckles. The fringe pattern is now a superposition of monochromatic fringe patterns. As the fringe spacing is wavelength-dependent, the resulting pattern becomes blurred far from the central fringe (Fig. \ref{Fig_SCC_poly_before_correction}), and the larger the bandwidth, the faster the fringes become blurred.

        \begin{figure}
        \centering
        \includegraphics[width = 0.36 \textwidth]{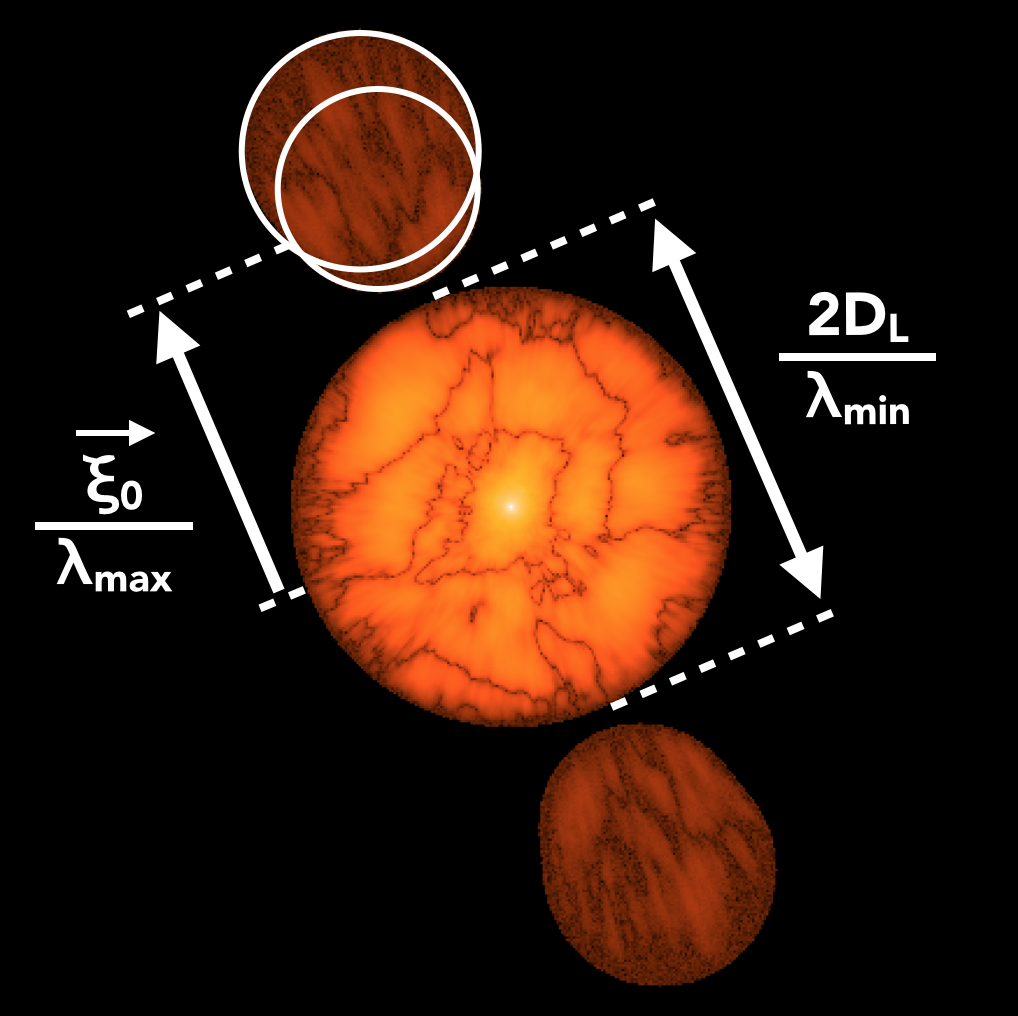} 
        \caption{Fourier transform of the SCC image obtained by numerical simulations, assuming polychromatic light ($\Delta\lambda=80\:\textrm{nm}$ at $\lambda_{0} = 640\:\textrm{nm}$).}
        \label{Fig_SCC_poly_FFT}
        \end{figure}

Figure \ref{Fig_SCC_poly_FFT} is the Fourier transform of the SCC image obtained in polychromatic light for $\Delta\lambda = 80\:\textrm{nm}$ around $\lambda_{0} = $ 640 nm. The associated equation of this Fourier transform is: 

\begin{eqnarray}
\mathcal{F}^{-1} \left[I_{\Delta\lambda}\right](\overrightarrow{u}) &=&  \int_{\Delta\lambda} \mathcal{F}^{-1} \left[\vert A_{S} \vert ^{2} + \vert A_{C}\vert ^{2} + \vert A_{R}\vert ^{2} \right] * \delta(\overrightarrow{u})\:d\lambda \nonumber \\
&+&  \int_{\Delta\lambda} \mathcal{F}^{-1} \left[A_{S}^{*}A_{R}\right] * \delta \left(\overrightarrow{u} -  \frac{\overrightarrow{\xi_{0}} }{\lambda} \right)d\lambda \nonumber \\   
&+& \int_{\Delta\lambda}  \mathcal{F}^{-1} \left[A_{S}A_{R}^{*}\right] * \delta \left(\overrightarrow{u} +  \frac{\overrightarrow{\xi_{0}} }{\lambda} \right)d\lambda .
\label{FT_SCC_poly}
\end{eqnarray}

The first term, corresponding to the central peak, now has a radius of $D_{L}/\lambda_{min}$. This is determined by the smallest wavelength $\lambda_{min} = \lambda_{0}-\Delta\lambda/2$. We still assume $\gamma \gg 1$. Each lateral peak is the sum of all the monochromatic lateral peaks. Their size and distance to the central peak are a function of the wavelength. The closest of these to the central peak has a radius $D_{L}(1 + 1/\gamma)/(2\lambda_{max})$ with $\lambda_{max} = \lambda_{0}+\Delta\lambda/2$. Thus, in polychromatic light, the lateral and central peaks do not overlap if and only if
\begin{eqnarray}
\vert\vert\overrightarrow{\xi_{0}} \vert\vert &>& 
D_{L}\left( \frac{\lambda_{max}}{\lambda_{min}} + \frac{1}{2} + \frac{1}{2\gamma}\right) .
\label{Eq_distance_poly}
\end{eqnarray}
If the condition of Equation \ref{Eq_distance_poly} is fulfilled, we can select the elongated lateral peak by using a circular oversized mask and extract $I_{-}$ using the procedure described for the monochromatic case (see Section \ref{Sec_SCC_sub_mono}). 

\subsection{Performance of the SCC in polychromatic light}
\label{subsec_SCCpolyperf}

\begin{figure}
        \centering
        \begin{subfigure}[b]{0.36\textwidth}      
                \includegraphics[width = \textwidth]{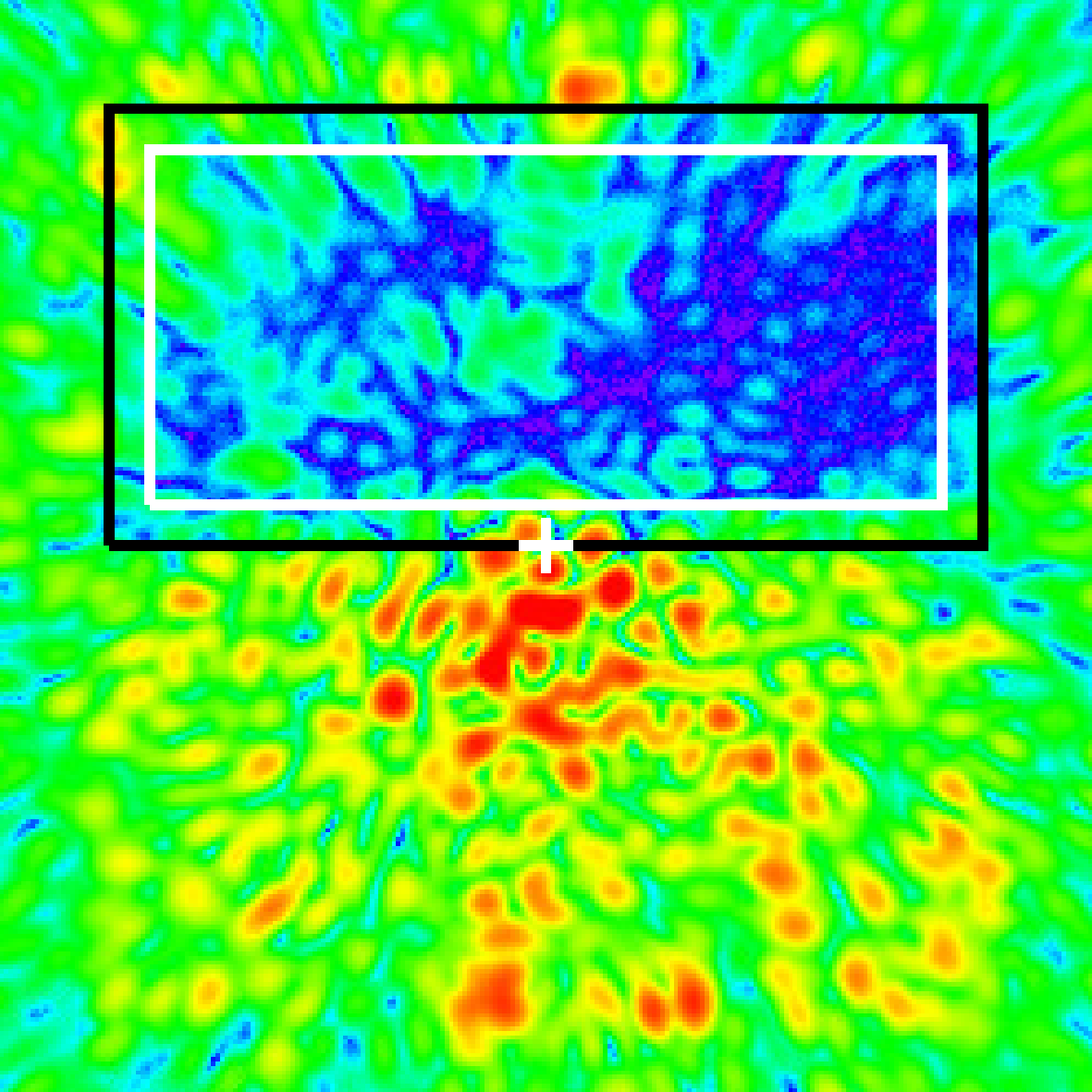}                                
        \end{subfigure}

        \begin{subfigure}[b]{0.36\textwidth}     
                \includegraphics[width = \textwidth]{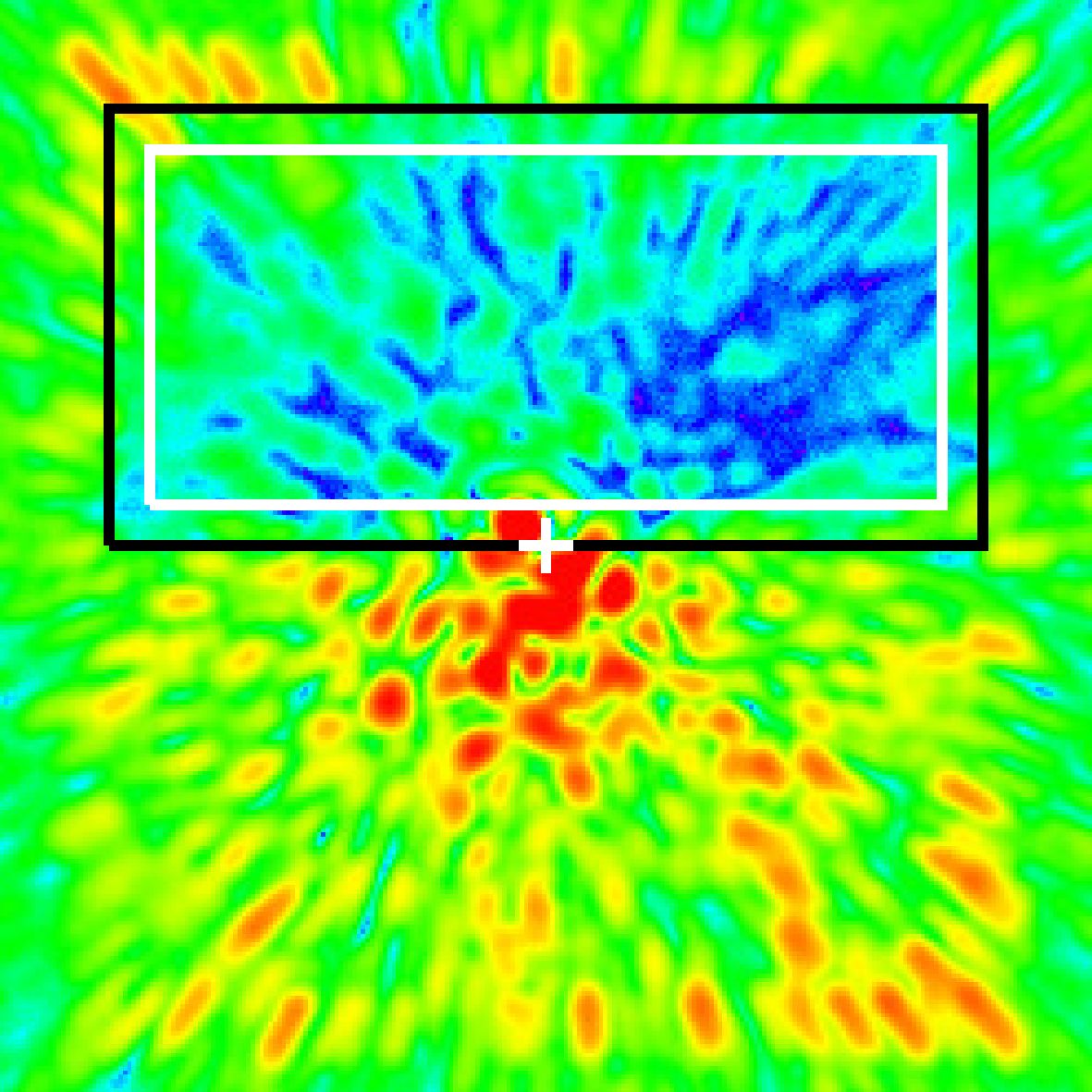}
        \end{subfigure}
           \caption{Numerically simulated coronagraphic images after correction with the SCC for $\Delta\lambda = 40\:\textrm{nm}$ (top) and $\Delta\lambda = 80\:\textrm{nm}$ (bottom).}
        \label{Fig_SCC_poly_with_readout_noise}%
\end{figure}

In this section we use the SCC in polychromatic light (Eq.\ref{Eq_distance_poly} fulfilled) to create a DH as performed in monochromatic light (Section \ref{Sec_SCC_sub_perf_mono}). We assume the parameters of Table \ref{Tab_B} and we obtain the image of Fig. \ref{Fig_SCC_poly_with_readout_noise} for a bandwidth of $40\:\textrm{nm}$ (top) and for $\Delta\lambda = 80\:\textrm{nm}$ (bottom). In both cases, $\vert A_{R} \vert ^{2}$ have the same characteristics as in monochromatic light (maximum at $4.8\:10^{-8}$ in normalized intensity).

Figure \ref{Fig_SCC_poly_with_readout_noise} highlights the limitation of the SCC in polychromatic light. When increasing the bandwidth, the SCC estimation is less and less efficient and speckles inside the DH are measured and are therefore not corrected (top left of the DH). Comparing these images with the images obtained before correction (Fig. \ref{Fig_SCC_poly_before_correction}), we observe that in the area where the fringes are blurred, the SCC estimation is disturbed. In other words, if speckles are not spatially modulated, the associated complex field $A_{S}$ cannot be estimated and thus, it cannot be minimized. As a consequence, the area of the DH where the correction is efficient quickly decreases as the bandwidth increases.

The cumulative functions associated with Fig. \ref{Fig_SCC_poly_with_readout_noise} are given in Fig. \ref{Fig_results_scc_mono_PC} (top). The performance degrades by a factor of three for the $\Delta\lambda = 40\:\textrm{nm}$ ($7.8\:10^{-8}$) and six for $\Delta\lambda = 80\:\textrm{nm}$ ($1.7\:10^{-7}$). These values confirm that the larger the bandwidth, the worse the correction.

To overcome the limitation of the SCC in polychromatic light, 
\cite{Galicher2008SPIE} proposed two methods: working with an integral field spectrometer (IFS) at modest spectral resolution (R = 30 to 100) or faking the use of a short bandpass filter using a Wynne compensator \citep{Wynne1979}. However, both methods have drawbacks that led us to develop new methods to make the SCC more achromatic. The Wynne compensator is a very invasive device composed of two triplets of lenses and the IFS is not available in all current coronagraphic instruments.

In this paper, we propose two other solutions to work with polychromatic light. The first method presented in Sect. \ref{Sec_piston} is based on the introduction of a non-null OPD between the reference hole and the classical Lyot stop. The second method, called the multireference self-coherent camera, uses several reference holes, instead of just one (Sect. \ref{Sec_MRSCC}).

\section{Optical path difference between reference hole and Lyot stop}
\label{Sec_piston}

bearing in mind that we reduce the DH to half of the influence area of the DM, part of the fringe pattern is located in the half plane that we do not try to correct. One solution to improve the performance of the SCC in polychromatic light consists of shifting the pattern of fringes inside the DH. To do so, we can introduce an achromatic OPD between the beam that goes through the classical Lyot stop and the beam that goes through the reference hole. Calling $\delta$ the OPD, Eq. \ref{Im_SCC_poly} becomes:

\begin{eqnarray} 
I_{\Delta\lambda}(\overrightarrow{\alpha}) &=&  \int_{\Delta\lambda}\left[ \vert A_{S} \vert ^{2} + \vert A_{C}\vert ^{2} + \vert A_{R}\vert ^{2}\right]d\lambda \nonumber \\
&+& \int_{\Delta\lambda} 2Re \left[A_{S}A_{R}^{*} \exp\left( \frac{2 i \pi \left(\overrightarrow{\alpha} . \overrightarrow{\xi_{0}} + \delta\right)}{\lambda}\right)\right]d\lambda .
\label{Eq_SCC_poly_piston}
\end{eqnarray}

\begin{figure}
        \centering
        \begin{subfigure}[b]{0.36\textwidth}
                \includegraphics[width = \textwidth]{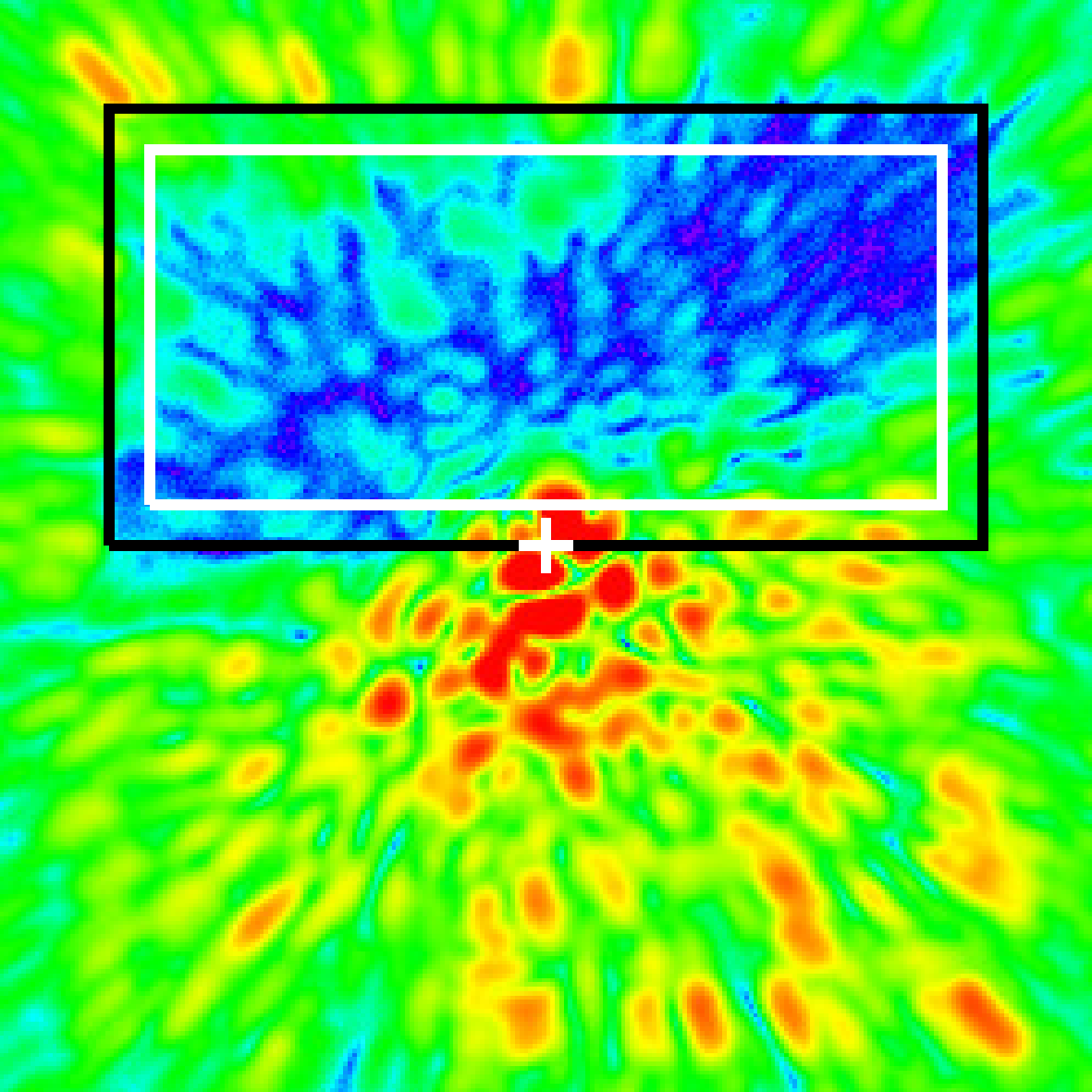}
        \end{subfigure}
        
        \begin{subfigure}[b]{0.36\textwidth}
                \includegraphics[width = \textwidth]{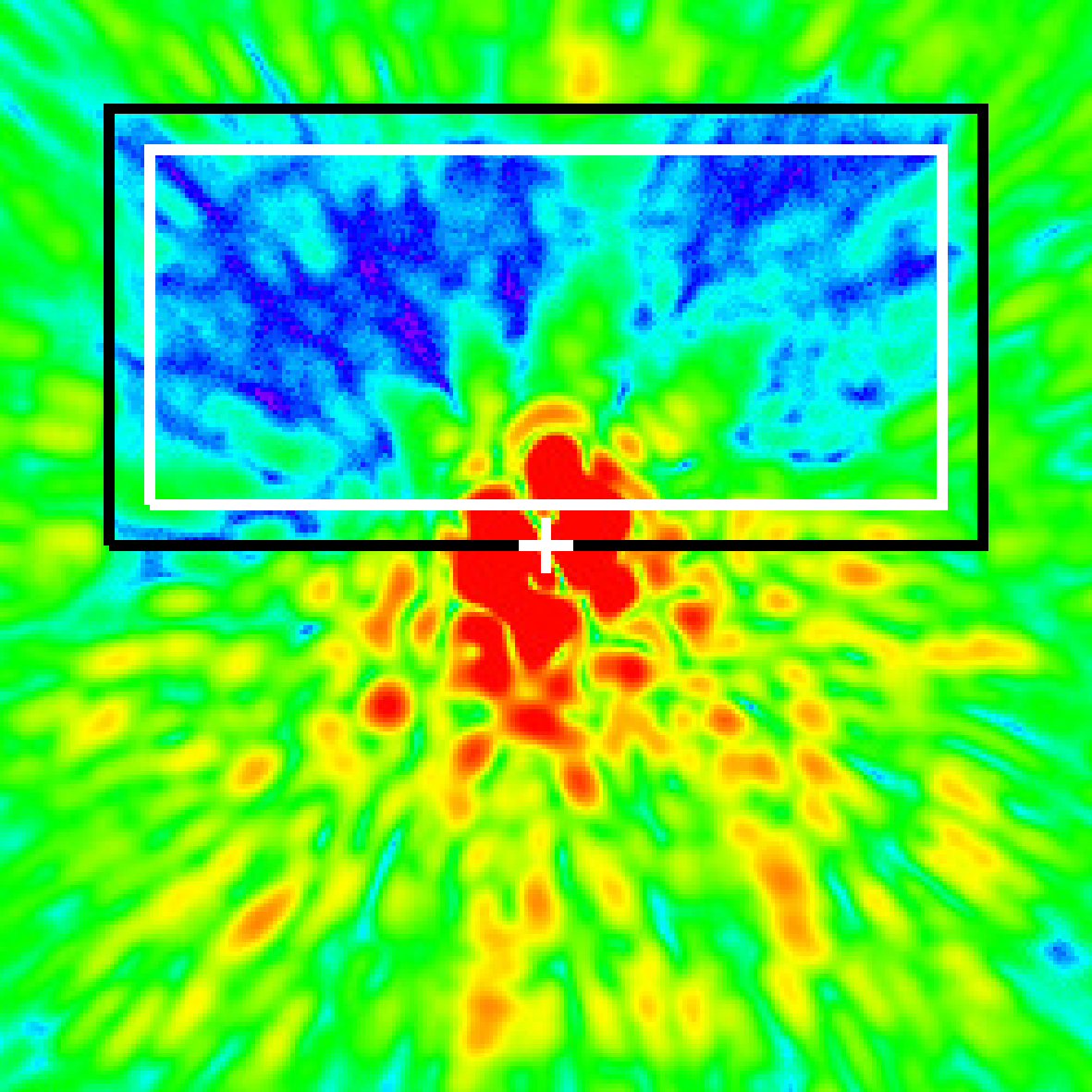}        
        \end{subfigure}
        \caption{Numerically simulated coronagraphic images for different OPD: (top) $\delta = 8\lambda_{0}$ and (bottom) $\delta = 16\lambda_{0}$. Bandwidth: $\Delta\lambda = $ 80 nm.}
        \label{Fig_Piston}
\end{figure}

When applying a constant OPD in the conditions of Fig. \ref{Fig_SCC_poly_with_readout_noise} (bottom) the fringe pattern can be shifted inside the DH. This increase of the fringe visibility inside the DH should improve the SCC estimation. Introducing an OPD of $\delta = 8\lambda_{0}$ or $\delta = 16\lambda_{0}$ shifts the central fringe by 8 or 16 fringes which corresponds to \textasciitilde $4.4\lambda_{0}/D$ and \textasciitilde $8.9\lambda_{0}/D$ in our numerical simulation (1.8 fringes per speckle). We obtain images of Fig. \ref{Fig_Piston} using the assumptions of Table \ref{Tab_B} and $\Delta\lambda = $ 80 nm.

\begin{figure}
        \centering
        \begin{subfigure}[b]{0.45\textwidth}
        \includegraphics[trim = 11mm 6mm -10mm 8mm, clip, width = \textwidth]{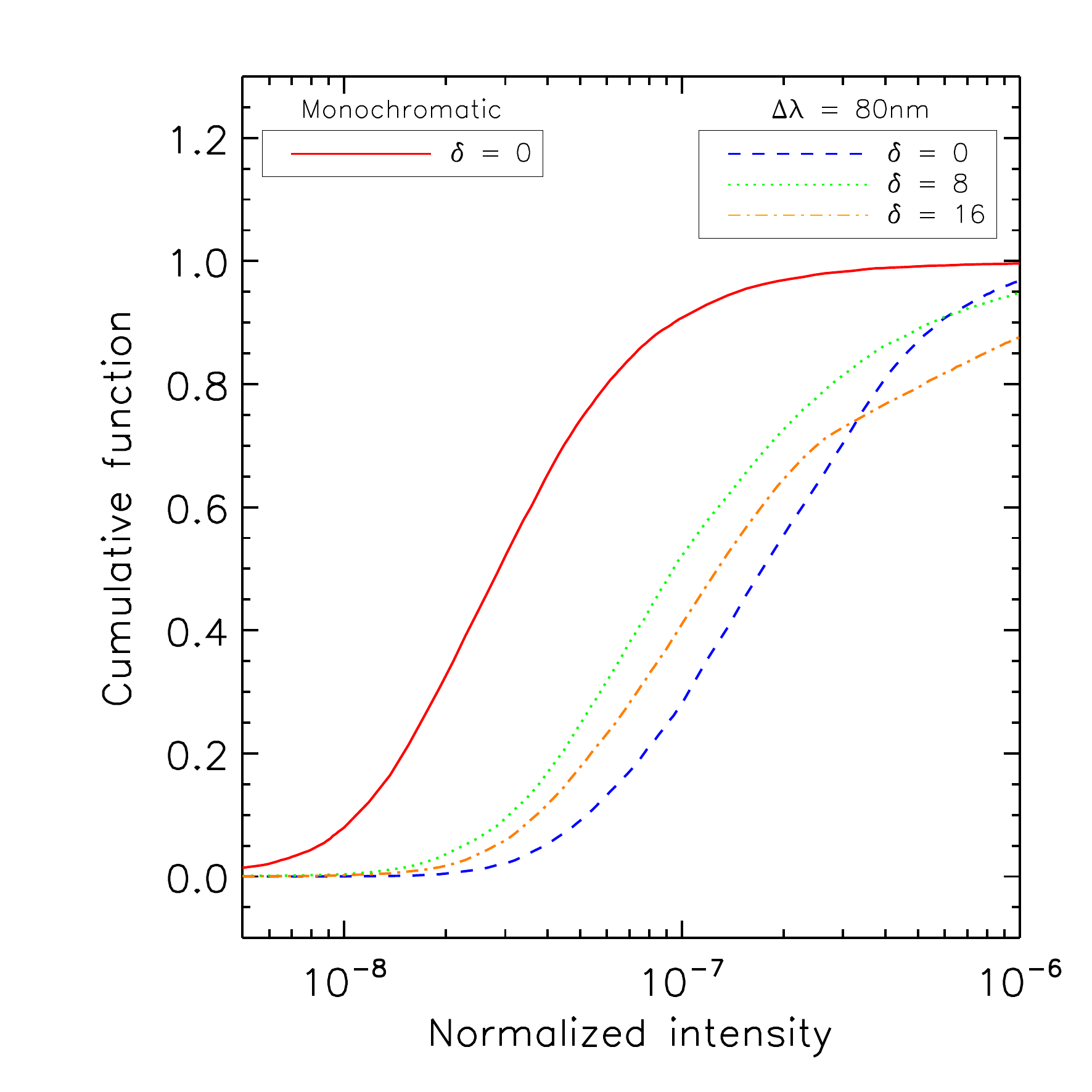} \\                         
        \end{subfigure}
        \caption{Cumulative functions associated with the images obtained by numerical simulations in monochromatic light (red solid line) and in polychromatic light ($\Delta\lambda = 80\:\textrm{nm}$ at 640 nm) assuming $\delta = 0$ (blue dashed line), $\delta = 8\lambda_{0}$ (green dotted line) or $\delta = 16\lambda_{0}$ (orange mixed line)}
        \label{Fig_piston_cumulative}
\end{figure}

Compared with Fig. \ref{Fig_SCC_poly_with_readout_noise}, we find that non-null OPD improves the SCC correction. Figure \ref{Fig_piston_cumulative} presents the cumulative curves associated with the images of Figs \ref{fig:mono_HDH:Im_coro}, \ref{Fig_SCC_poly_with_readout_noise} (bottom), and \ref{Fig_Piston}.  In the case with no OPD, the median of the speckle intensity is $1.7\:10^{-7}$, while we find $9.4\:10^{-8}$ and $1.4\:10^{-7}$ for $\delta = 8\lambda_{0}$ and $\delta = 16\lambda_{0}$. We find the  $\delta = 16\lambda_{0}$ case is less efficient than the $\delta = 8\lambda_{0}$ case. Indeed, even if the fringe pattern is in the middle of the DH when $\delta = 16\lambda_{0}$, the visibility of the SCC fringes is not uniform in the DH. Visibility is low close to the optical axis because the speckles are a lot brighter than the $\vert A_{R} \vert ^{2}$ there. That is why the central part of the DH is not well corrected in the $\delta = 16\lambda_{0}$ case. The best value of the OPD is when the white fringe is closer to the axis: a smaller OPD.

Even if the OPD solution can improve the contrast inside the DH, we noticed in our numerical simulations that the performance strongly depends on the initial speckles pattern and when low order aberrations are too large, the correction does not converge. This technique would then be difficult to use and we propose another solution to improve the performance of the SCC estimation and correction: the MRSCC.

\section{Multireference self-coherent camera}
\label{Sec_MRSCC}

The MRSCC is designed to mitigate the chromatic limitation of the SCC. In this section we introduce the formalism of the MRSCC and study the performance as a function of the bandwidth from numerical simulations.

\subsection{Formalism of the MRSCC}
\label{SubSec_MRSCCFormalism}

\begin{figure}
        \centering
        \begin{subfigure}[b]{0.3\textwidth}
                \includegraphics[width = \textwidth]{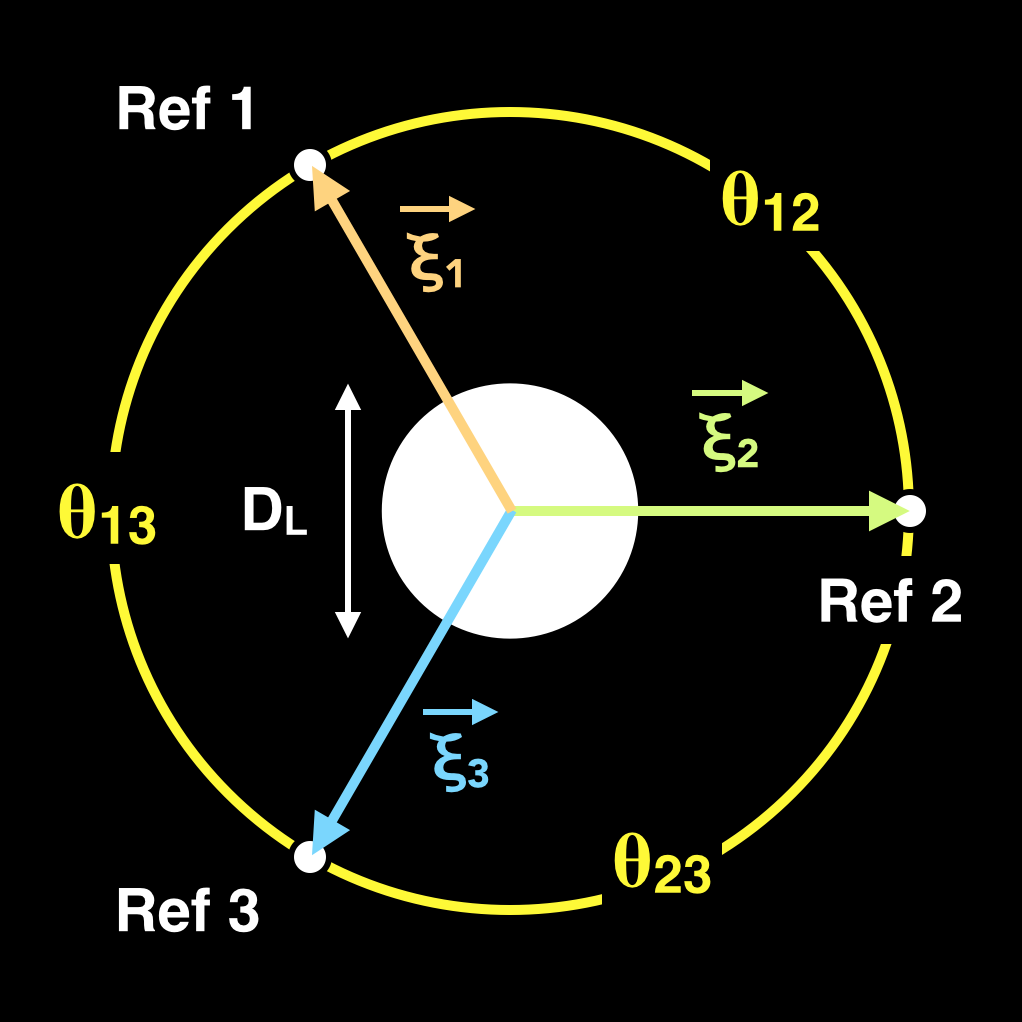}
        \end{subfigure}

        \begin{subfigure}[b]{0.3\textwidth}
          \includegraphics[width = \textwidth]{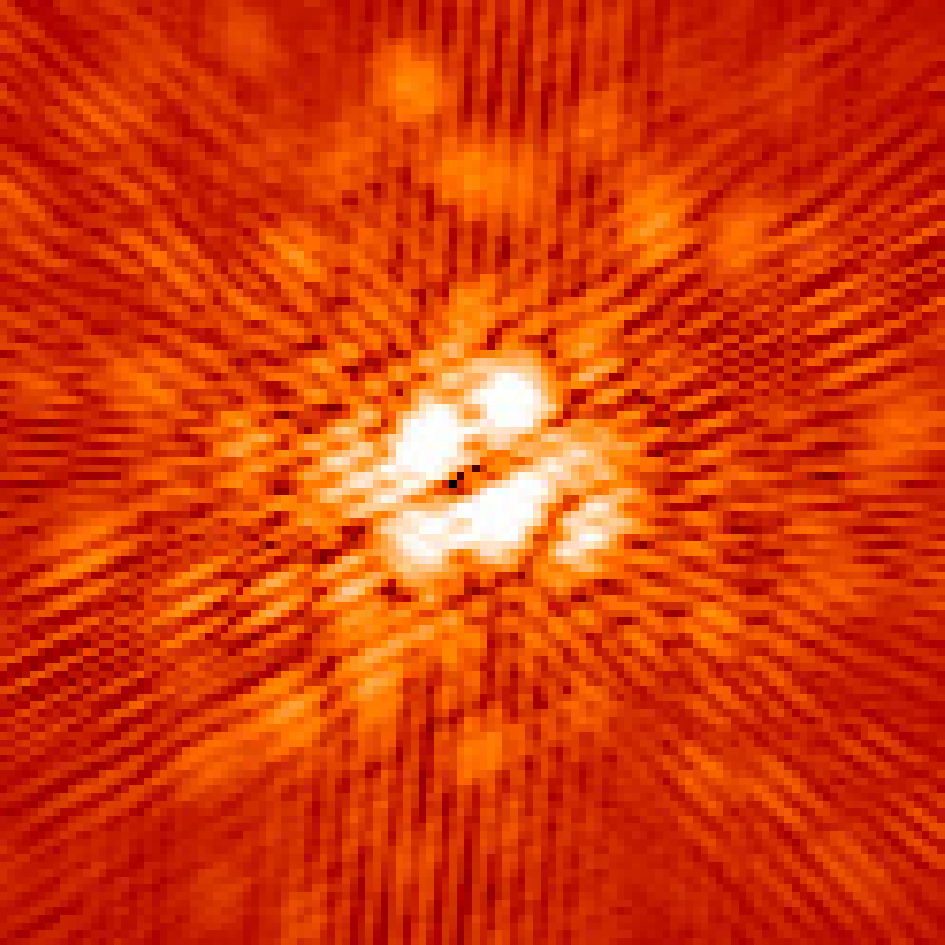}
        \end{subfigure}

        \begin{subfigure}[b]{0.3\textwidth}      
        \includegraphics[width = \textwidth]{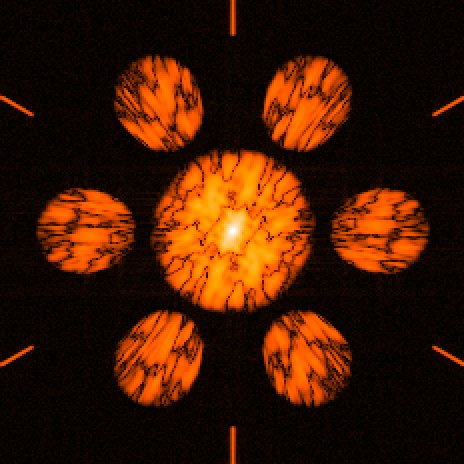}
        \end{subfigure}
        \caption{Top: MRSCC Lyot stop with three reference holes. Middle: Numerically simulated MRSCC image showing the three fringe patterns before correction. Field of view: 20 by 20 $\lambda_{0}/D$. Bottom: Fourier transform of the MRSCC image.} 
        \label{Fig_EEZ}
\end{figure}

The fringe pattern produced by the reference hole of the SCC gets blurred far from the central fringe in polychromatic light. In these blurred areas, the speckles cannot be corrected. To increase the area of the DH where the speckles are spatially modulated, we introduce additional fringe patterns by adding reference holes in the Lyot stop (Fig. \ref{Fig_EEZ}, top). As for the SCC reference hole, the stellar light selected by the new reference holes interferes with the stellar coronagraphic residue in the image plane, which creates Fizeau fringes on the speckles. The orientation and the fringe spacing of these new patterns depend on the positions of the reference holes with respect to the Lyot stop. By placing them shrewdly it is possible to spatially modulate areas of the DH where the fringes that are induced by the SCC hole are blurred. Thus, the MRSCC can increase the area where speckles are correctly modulated.

Figure \ref{Fig_EEZ} (middle) presents one MRSCC image obtained with a Lyot stop that was composed of three reference holes in addition to the classical Lyot stop (Fig. \ref{Fig_EEZ}, top). In this MRSCC image, we numerically increased the flux into the reference holes to highlight 
the fringes. The distances and the diameter ratio between the reference holes and the Lyot stop pupil are the same ($\xi_{1} = \xi_{2} = \xi_{3} = 1.8D_{L}$ and $\gamma_{1} = \gamma_{2} = \gamma_{3} = 25$). They obey Eq. \ref{Eq_distance_poly} for $\Delta\lambda = 80\:\textrm{nm}$. The angle between each couple of reference holes is $120^{\circ}$. We can write the intensity in the MRSCC image as

\begin{eqnarray}
I_{\Delta\lambda}(\overrightarrow{\alpha}) &=&  \int_{\Delta\lambda}\left[\vert A_{S} \vert ^{2} + \vert A_{C}\vert ^{2} + \sum\limits_{i=1}^{n} \vert A_{R_{i}}\vert ^{2} \right] d\lambda \nonumber \\
&+&  \sum\limits_{i=1}^{n} \int_{\Delta\lambda} 2Re \left[ A_{S}A_{R_{i}}^{*} \exp\left( \frac{2 i \pi \overrightarrow{\alpha} . \overrightarrow{\xi_{i}}}{\lambda}\right)\right] d\lambda  \nonumber  \\
&+&  \sum\limits_{j=1}^{n} \sum\limits_{i > j}^{n} \int_{\Delta\lambda}2Re \left[ A_{R_{j}}A_{R_{i}}^{*}\exp\left( \frac{2 i \pi \overrightarrow{\alpha} . \overrightarrow{\xi_{ij}}}{\lambda}\right)\right] d\lambda ,
\label{Eq_SCC_poly_multi}
\end{eqnarray}
where n is the number of reference holes and $\overrightarrow{\xi_{ij}} = \overrightarrow{\xi_{i}} - \overrightarrow{\xi_{j}}$. 
The second term is the sum of the modulations of the speckle electric field $A_{S}$ by each reference beam $A^{*}_{R_{i}}$. The third term contains the sum of the fringe patterns that are due to the interference between reference holes. Indeed, beams coming from two reference holes also form Fizeau fringes in the MRSCC image. To order to estimate the complex amplitude $A_{S}$ we apply the Fourier transform to the image (Fig.\ref{Fig_EEZ}, bottom): 

$\mathcal{F}^{-1} \left[I_{\Delta\lambda}\right](\overrightarrow{u}) = $
\begin{eqnarray}
&& \int_{\Delta\lambda} \left( \mathcal{F}^{-1} \left[\vert A_{S} \vert ^{2} + \vert A_{C}\vert ^{2}\right]  + \sum\limits_{i=1}^{n} \mathcal{F}^{-1} \left[ \vert A_{R_{i}}\vert ^{2}\right]\right)*\delta \left( \overrightarrow{u}\right) d\lambda \nonumber \\
&+&  \int_{\Delta\lambda}  \sum\limits_{i=1}^{n} \mathcal{F}^{-1} \left[A_{S}^{*}A_{R_{i}}\right] * \delta \left( \overrightarrow{u} - \frac{\overrightarrow{\xi_{i}} }{\lambda}\right) d\lambda \nonumber \\   
&+& \int_{\Delta\lambda} \sum\limits_{i=1}^{n} \mathcal{F}^{-1} \left[A_{S}A_{R_{i}}^{*}\right] * \delta \left(\overrightarrow{u} + \frac{\overrightarrow{\xi_{i}} }{\lambda} \right)d\lambda \nonumber \\   
&+&  \int_{\Delta\lambda}  \sum\limits_{j=1}^{n} \sum\limits_{\substack{i=1 \\ i\neq j} }^{n} \mathcal{F}^{-1} \left[A_{R_{j}}^{*}A_{R_{i}}\right] * \delta \left( \overrightarrow{u} -  \frac{\overrightarrow{\xi_{ij}} }{\lambda} \right) d\lambda .
\label{FT_SCC_mono_multi}
\end{eqnarray}
The first term is the central peak of the spatial frequency plane. The second and third term contain the lateral correlation peaks of $A_{S}$ and  $A^{*}_{R_{i}}$ and their conjugate. There are a couple of conjugated correlation peaks per reference hole. This is the information we want to extract to control the DM and create the DH. Finally, the last term contains the correlation peaks between reference holes. They do not disturb the estimation of the electric field if they do not overlap the correlation peaks used to do the estimation of $I_{-}$ (see Fig. \ref{Fig_EEZ} bottom, six small sections of straight line at the border of the Fourier image). The size and the position of each correlation peak depend on the spectral bandwidth and on the position of the reference holes in the Lyot stop plane.

To estimate the electric field, we extract each lateral peak of the second term. Then, using the algorithm described in Sect. \ref{Sec_SCC_sub_mono} we obtain one $I_{-}$ for each reference hole. In polychromatic light, these estimations are different and complementary because each reference beam modulates different speckles (i.e., different parts of the focal plane image). We build a new polychromatic estimator which is the concatenation of the n estimators provided by the n reference beams. Using this new estimation of the electric field is the only change from the SCC correction loop described in Sect. \ref{Sec_SCC_sub_mono}. 

For an efficient estimation of the speckle electric field, we ensure that the different components in the spatial frequency plane do not overlap. First, the distances between reference holes and the Lyot stop pupil obey Eq. \ref{Eq_distance_poly}. Then, the correlation peaks do not overlap. Finally, the position of each reference hole is chosen so that the secondary lateral correlation peaks do not overlap the primary correlation peaks. 

\subsection{Performance of the MRSCC}
\label{Sec_param}

In this section, we simulate an MRSCC with three reference holes assuming monochromatic light $\lambda_{0}=  640\:\textrm{nm}$ and two bandwidths ($\Delta\lambda = 40\:\textrm{nm}$ and $\Delta\lambda = 80\:\textrm{nm}$) around $\lambda_{0}=  640\:\textrm{nm}$. The angle $\theta_{H}$ between the first reference and the horizontal axis in the Lyot stop plane is $+115^{\circ}$. The angle between the first reference hole and the second one $\theta_{12}$ is $-135^{\circ}$ and there is an angle $\theta_{13}$ of $+105^{\circ}$ between the first reference hole and the third one. These angles correspond to the ones used in our laboratory experiment (see Sect. \ref{Sec_THD}). Separations between the classical Lyot stop and reference holes are: $\xi_i = 1.8 D_{L}$. We assume $\gamma_{i} = 25$. Because we simulate a perfect coronagraph, we numerically add the reference beams and we set their flux to the one measured in our laboratory. 
The maximum normalized intensity level of $\vert A_{R_{i}}\vert ^{2}$ is $4.8\:10^{-8}$. Finally, the other parameters are the same as previously (see Table \ref{Tab_B}). 

\begin{figure}
        \centering
        \begin{subfigure}[b]{0.36\textwidth} 
                \includegraphics[width = \textwidth]{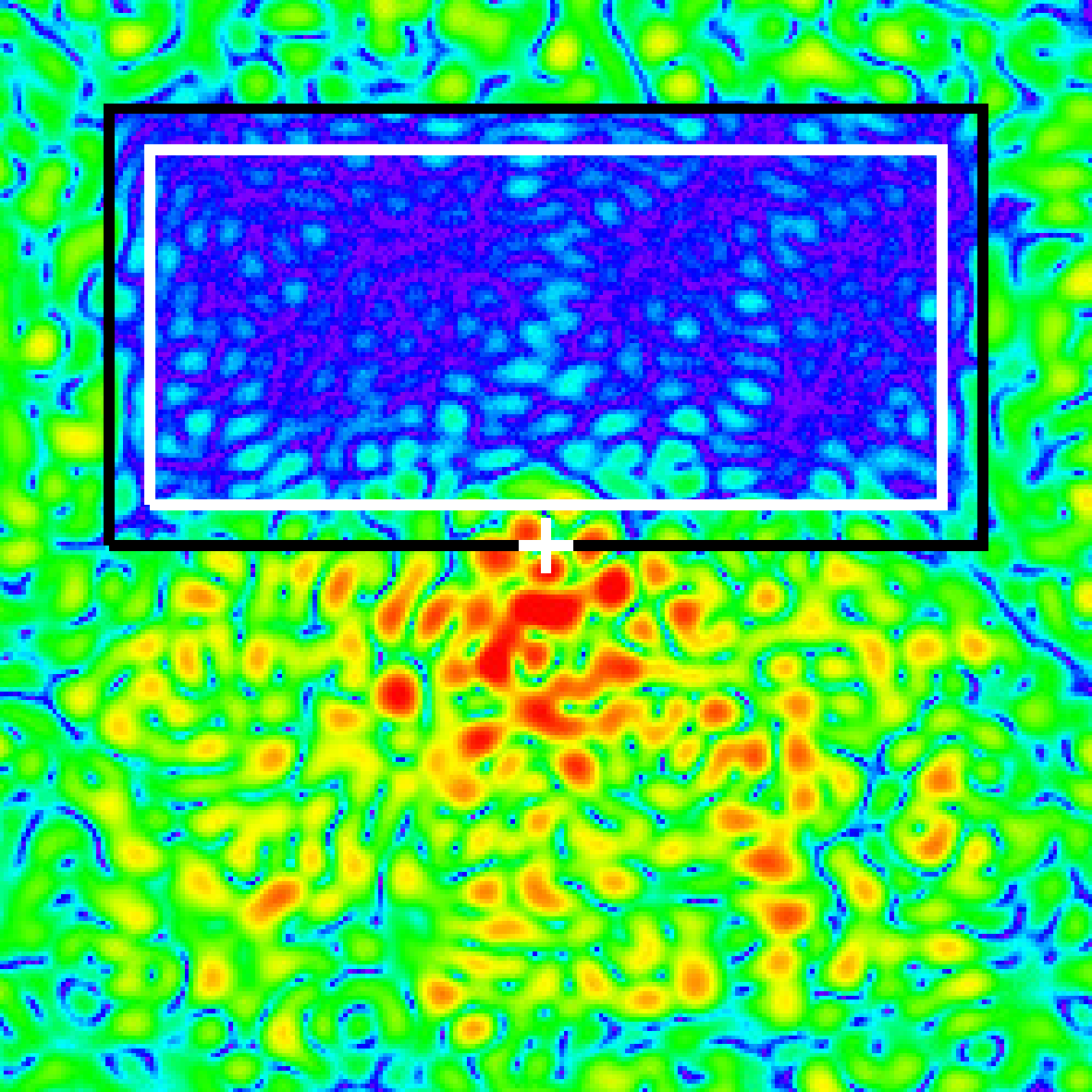}        
        \end{subfigure}
        
        \begin{subfigure}[b]{0.36\textwidth}               
                \includegraphics[width = \textwidth]{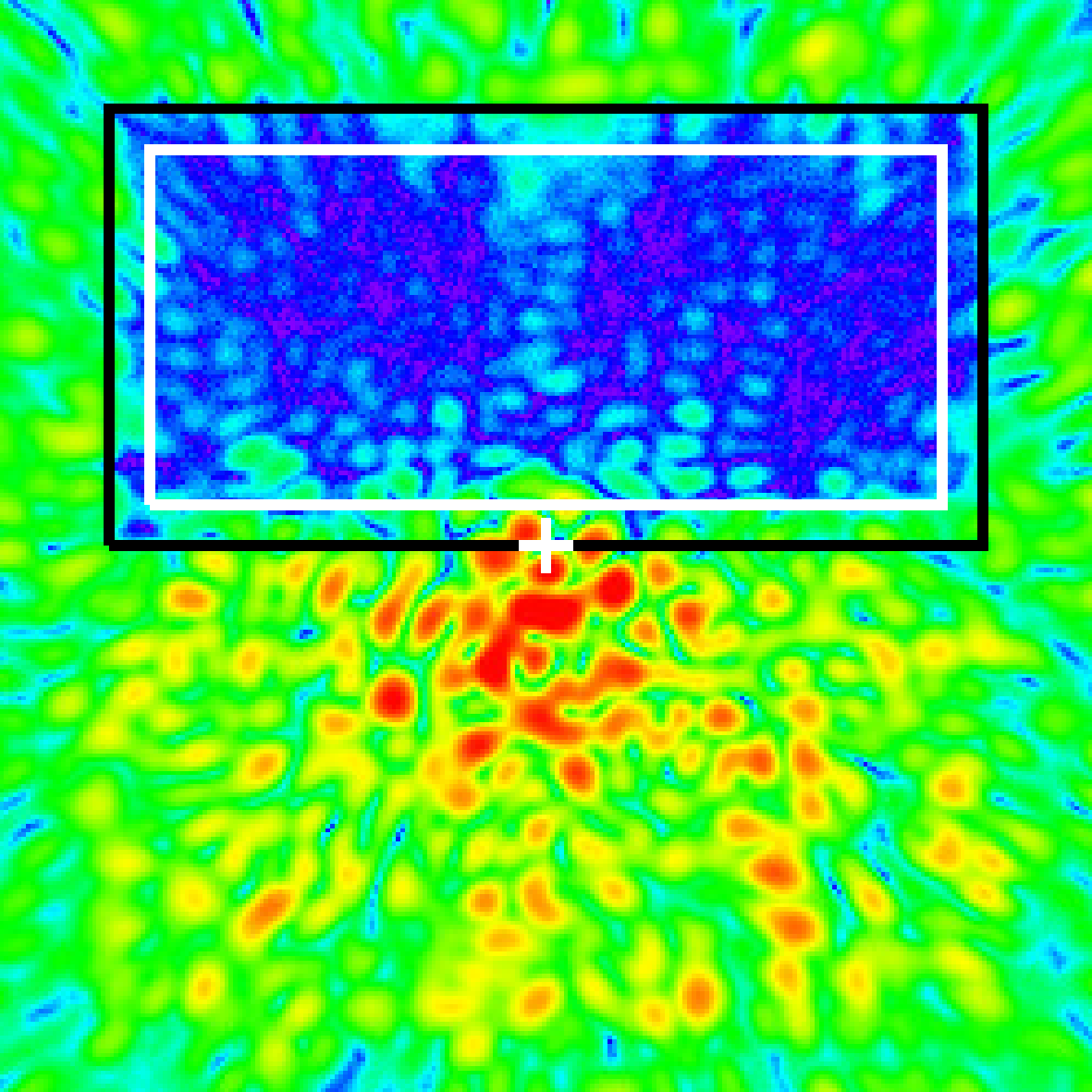}        
        \end{subfigure}

        \begin{subfigure}[b]{0.36\textwidth}
                \includegraphics[width = \textwidth]{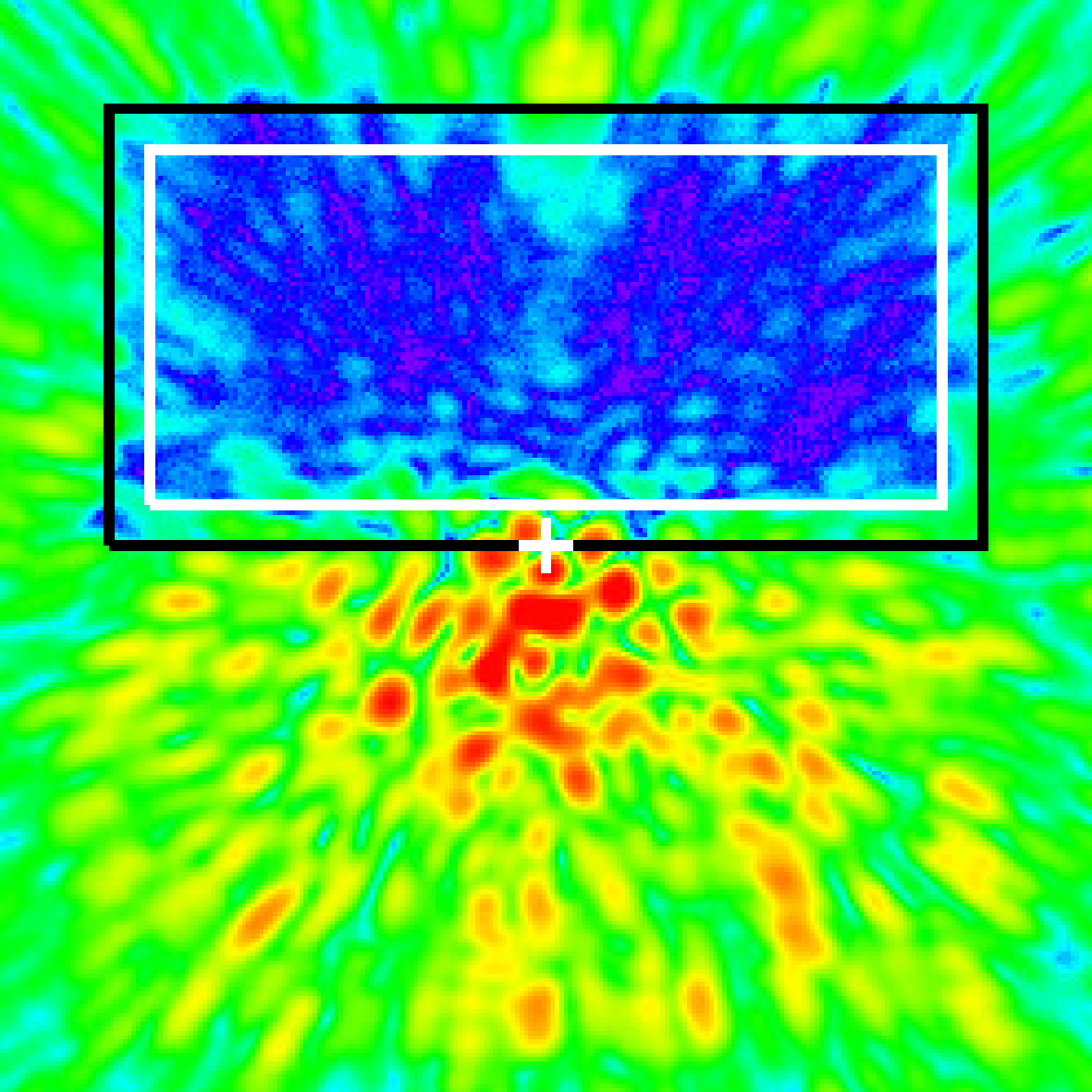}        
        \end{subfigure}
        \caption{Numerically simulated coronagraphic images obtained after correction with the MRSCC in monochromatic light (top) ($\lambda_{0}=  640\:\textrm{nm}$) and in polychromatic light ($\Delta\lambda = 40\:\textrm{nm}$ (middle) and $\Delta\lambda = 80\:\textrm{nm}$ (bottom) around $\lambda_{0}=  640\:\textrm{nm}$).} 
        \label{Fig_bandwidth_nbref}
\end{figure}

Figure \ref{Fig_bandwidth_nbref} presents the images after correction of the speckles using the MRSCC as a focal plane wavefront sensor. At the top, the image was obtained after correction in monochromatic light. If we compare this image and the image obtained in monochromatic light with the SCC (Figure \ref{fig:mono_HDH:Im_coro}), we can see that they are almost the same. Indeed, in monochromatic light, fringes do not get blurred and adding reference holes bring no new information on $A_{S}$.

The middle and bottom images were obtained after correction in polychromatic light: $\Delta\lambda = 40\:\textrm{nm}$ and $\Delta\lambda = 80\:\textrm{nm}$ around $\lambda_{0}=  640\:\textrm{nm}$. They show that, when increasing the bandwidth, speckles in the DH of the MRSCC images are well corrected for bandwidth up to 80 nm.

        \begin{figure}
        \centering
        \includegraphics[trim = 11mm 6mm -10mm 8mm, clip, width = 0.45 \textwidth]{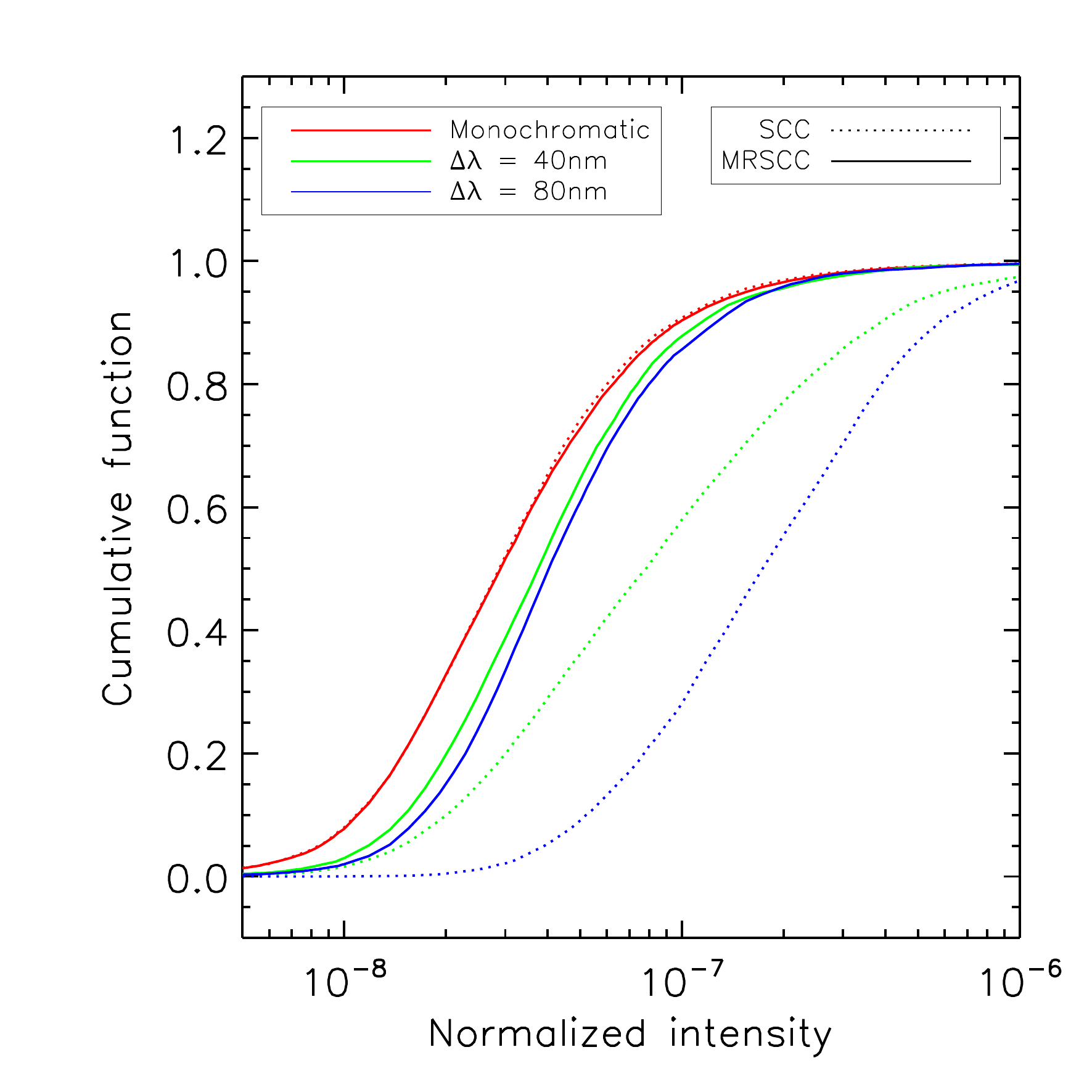} \\             
        \caption{Cumulative functions associated with the images obtained by numerical simulations with the SCC (dashed curve) and with the MRSCC (solid curves). In monochromatic light (red curves) the cumulatives functions of the both SCC and MRSCC are the same.}
        \label{Fig_C}
        \end{figure}

Figure \ref{Fig_C} presents the cumulative functions associated with these three images (solid lines) and to the images obtained with the SCC of Figs. \ref{fig:mono_HDH:Im_coro} and \ref{Fig_SCC_poly_with_readout_noise} (dotted lines). In monochromatic light, the performances are almost the same using one or three reference holes. The slight difference can be explained by the fact that photon noise is greater for the MRSCC (three reference beams) than for the SCC (one reference beam). In polychromatic light, MRSCC always provides images with deeper contrast levels because speckles that were not well estimated with the unique reference hole of the SCC are estimated using the second or the third reference hole of the MRSCC. The median of the speckle intensity inside the DH is $3.8\:10^{-8}$ for $\Delta\lambda = 40$ nm and $4.1\:10^{-8}$ for $\Delta\lambda = 80$ nm. We find a gain of \textasciitilde 2 for $\Delta\lambda = 40$ nm and \textasciitilde 4 for $\Delta\lambda = 80$ nm in contrast to the SCC performance (see Section \ref{subsec_SCCpolyperf}).
Thus, the MRSCC significantly mitigates the chromatic effects of the SCC and its performance is almost insensitive to the bandwidth up to $\Delta\lambda = 80$ nm. The small difference is the result of a few bright speckles located close to the border of the DH. 

Unlike the OPD method (Sect. \ref{Sec_piston}), the MRSCC performance does not depend on the initial map of phase and amplitude aberrations. It also reaches deeper contrast levels in numerical simulations. Of course MRSCC is not insensitive to chromatism. Its performance decreases gradually when the bandwidth is widened. However, from preliminary numerical simulations that are not presented in this paper, it can be used to control contrast levels in the DH by working with bandwidths up to 150nm in visible light. For larger bandwidths, we may add a fourth reference hole, bearing in mind the constraints presented in Sect. \ref{SubSec_MRSCCFormalism}. Another solution could be a combination of the OPD method although it may be complicated to achromatically control the OPD. In the following sections we probe the MRSCC performance in the laboratory. 

\section{The THD bench and expected performance}
\label{Sec_THD_description}
In this section we present our laboratory optical bench called the THD-bench. We use it to derive the expected performance of the MRSCC from numerical simulations.

\subsection{The THD Bench}
\label{SubSec_THD}

The SCC and MRSCC are tested in our laboratory at the Observatoire de Paris. A complete description of the optical bench is given in \cite{Mas2010}. Here, we only list main components:

\begin{itemize}
\item Two light sources. A quasi-monochromatic laser diode emitting at $\lambda_{0} = $ 637 nm ($\Delta\lambda < 1$ nm) and a supercontiniuum Fianium source (SC450) with a calibrated spectral filter ($\lambda_{0} = 640$ nm $\Delta\lambda = 80$ nm).
\item A tip-tilt mirror. Control of pointing errors down to $6.5\:10^{-2} \lambda_{0}/D$ \citep{Mas2012}.
\item A Boston Micromachines Corporation DM of 32 by 32 actuators in a square array. This is set in a pupil plane that is upstream of the coronagraph focal mask and we use only 27 actuators across the pupil diameter.
\item a dual-zone phase mask (DZPM) coronagraph composed of an apodizer in pupil plane and a phase mask in focal plane \citep{Soummer2003,NDiaye2012} that are optimized to reach  1$\sigma$  contrast under $2.8\:10^{-8}$ for an angular separation greater than 5 $\lambda_{0}/D$ and for a bandwidth of 133 nm around 665 nm.  

\item A Lyot stop (Fig. \ref{Fig_EEZ}, top) with a classical Lyot stop aperture with a $D_{L}$ = 8 mm diameter for a \O$_{p} $ =  8.1 mm entrance pupil (98.7\% filtering) and three reference holes (see Fig. \ref{Fig_EEZ} top:  $\theta_{H} = +115^{\circ}$, $\theta_{12}$ = -146$^{\circ}$, $\theta_{13}$ = +103$^{\circ}$, $\gamma_{1} = \gamma_{2} = \gamma_{3} = 26.6$, $\xi_{1}$ = 1.76$D_{L}$, $\xi_{2}$ = 1.89$D_{L}$ and $\xi_{3}$ = 1.93$D_{L}$).
\item An Andor camera with a readout noise of 3.2 e$^{-}$ RMS per pixel and a full well capacity of 60,000 e$^{-}$ per pixel. We used 400 by 400 pixel images and the resolution element $\lambda_{0}/D_{L}$ is sampled by 6.25 pixels.
\end{itemize}

The only component modified to switch from the SCC to the MRSCC is the Lyot stop. By not disturbing the optical path thanks to a motorised mount, we can open or close the reference holes. We note that we did not use the motor to produce a temporal modulation as \cite{giveon2012}. As with the SCC, the MRSCC uses spatial modulation and it only needs a single image to retrieve the complex amplitude of the speckle field.

\begin{figure}[h!]
        \centering
        \begin{subfigure}[b]{0.48\textwidth}     
        \includegraphics[width = \textwidth]{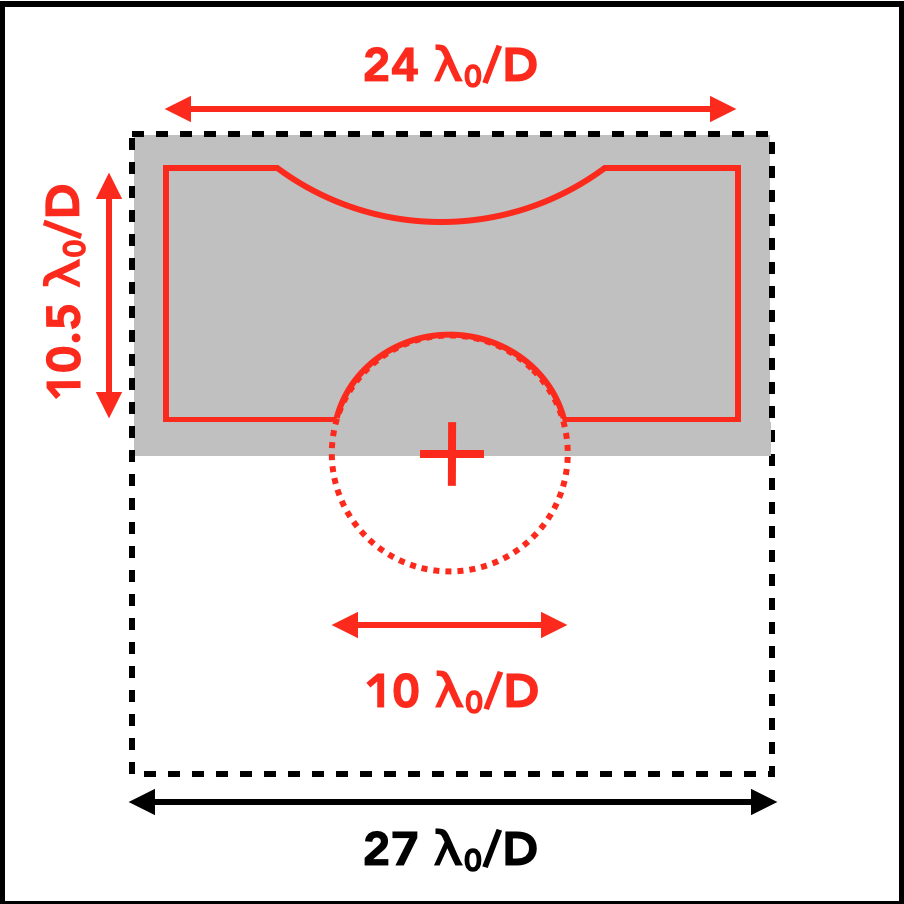}
        \label{Fig_THD_DH:1}
        \end{subfigure}
        \caption{Main dimensions used in our laboratory experiment. The influence area of the DM is delineated by a dashed line, the DH is represented by a gray area and the computation area is delineated by a red solid line. The optical axis is represented by the red cross.}
        \label{Fig_THD_DH}
\end{figure}

\begin{figure}[h!]        
        \centering
        \begin{subfigure}[b]{0.35\textwidth}
        \includegraphics[trim = 11mm 6mm -10mm 8mm, clip, width = 1.1 \textwidth] {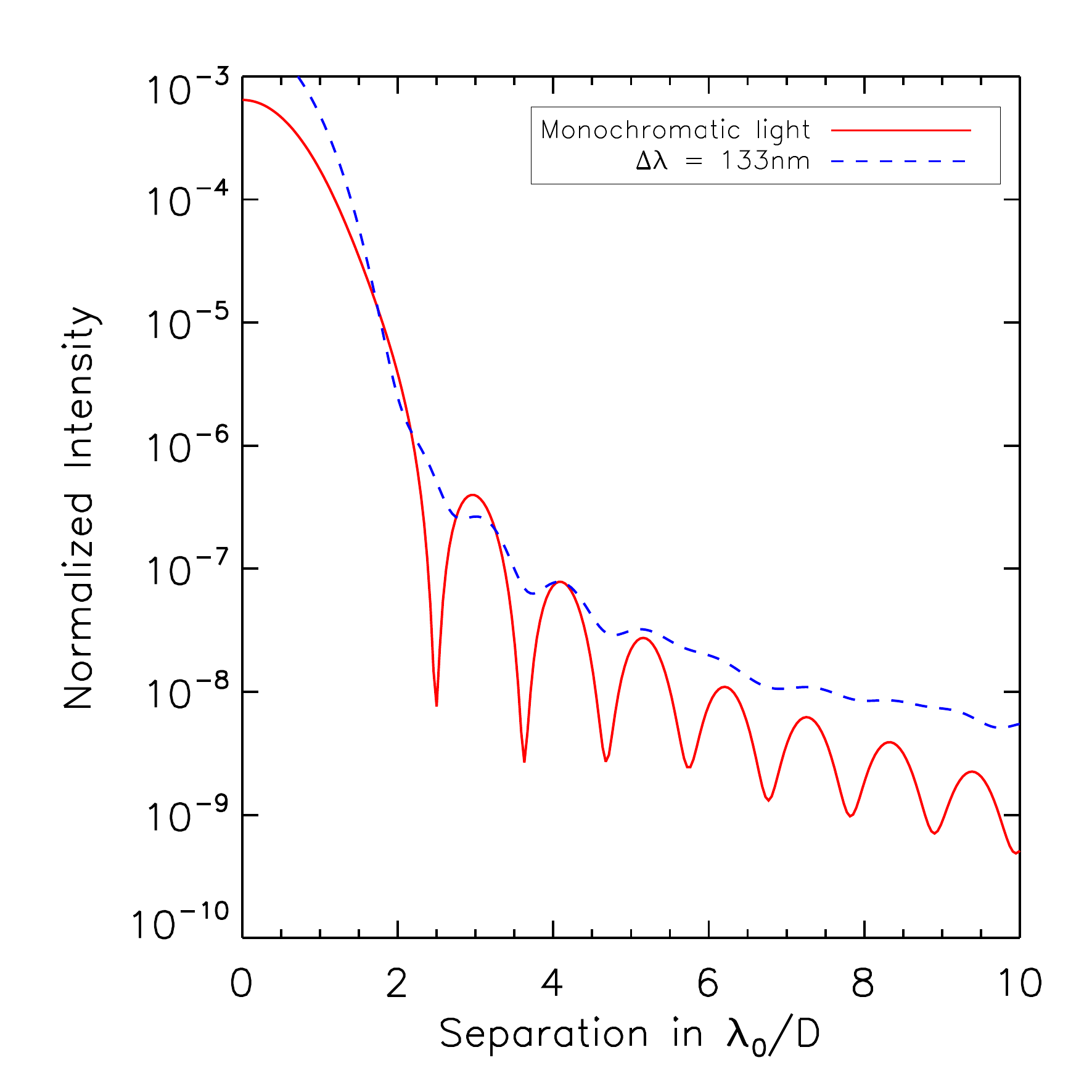}
        \end{subfigure}
        \caption{Radial intensity profiles computed from coronagraphic images obtained by numerical simulations without phase and amplitude aberrations. The red solid curve is the profile associated with an image obtained in monochromatic light while, the blue dashed lines are obtained in polychromatic light ($\Delta\lambda =  133$ nm around $\lambda_{0} =  665$ nm).}
        \label{referee_C}
\end{figure}

With 27 actuators across the pupil diameter, we can control a 13.5 by 27 $\lambda_{0}/D$ DH area with the use of only one DM to correct the effects of both phase and amplitude aberrations (Fig. \ref{Fig_THD_DH}). Even in an ideal case (without aberrations), the DZPM coronagraph does not perfectly cancel the star. In Fig. \ref{referee_C}, the red curve represents the theorerical limit of the DZPM coronagraph in monochromatic light ($\lambda_{0} = 640$ nm) and the blue dashed curve shows a bandwidth of $\Delta\lambda = 133$ nm around $\lambda_{0} = 665$ nm. Because of the design specifications, most of the speckles below 5 $\lambda_{0}/D$ are very bright and some saturate the detector. To prevent the SCC estimation being biased, we numerically reduce the speckles' impact by multiplying the SCC images by a Butterworth-type spatial filter, shaped like a disk centered on the optical axis. This numerical mask multiplies the half-DH Butterworth mask and the resulting mask attenuates the central speckles by a factor of 3 and those located at 5 $\lambda_{0}/D$ by a factor 1.5. The computation area is represented by a solid red line. To obtain this area, we remove a band of 1.5$\lambda_{0}/D$ on each side of the DH. To avoid the bright speckles, which result from the DZPM, we remove a disk centered on the optical axis of $<10\lambda_{0}/D$ diameter and because of bright speckles at the top of the DH (Sect. \ref{Sec_THD}) we also remove a disk centered at 22 $\lambda_{0}/D$ from the optical axis and with a radius of 12.5$\lambda_{0}/D$. 

\subsection{Numerical simulations of the THD bench}
\label{SubSec_TR}

In this section, we run numerical simulations of the MRSCC in polychromatic light ($\lambda_{0}=640\:\textrm{nm}$ - $\Delta\lambda=80\:\textrm{nm}$) using the parameters that best reproduce the optical configuration of the THD bench (see Section \ref{SubSec_THD}). Amplitude aberrations are derived from a laboratory measurement \citep{Mazoyer2013a}. Phase aberrations have a standard deviation of 10 nm and a power spectral density of $f^{-2.3}$, which best reproduce the speckles' intensity distribution in non-corrected images. We numerically simulate the electric field produced by the DZPM coronagraph inside the classical Lyot stop, as well as inside each reference hole. We find that the flux inside the reference beams are similar to the flux measured on the THD bench. Finally we use the measured spectrum of the 80 nm source (close to a flat spectrum) that was used during the experiments (Sect. \ref{Sec_THD}).
\begin{figure}
        \centering
        \begin{subfigure}[b]{0.36\textwidth}
        \includegraphics[width = \textwidth]{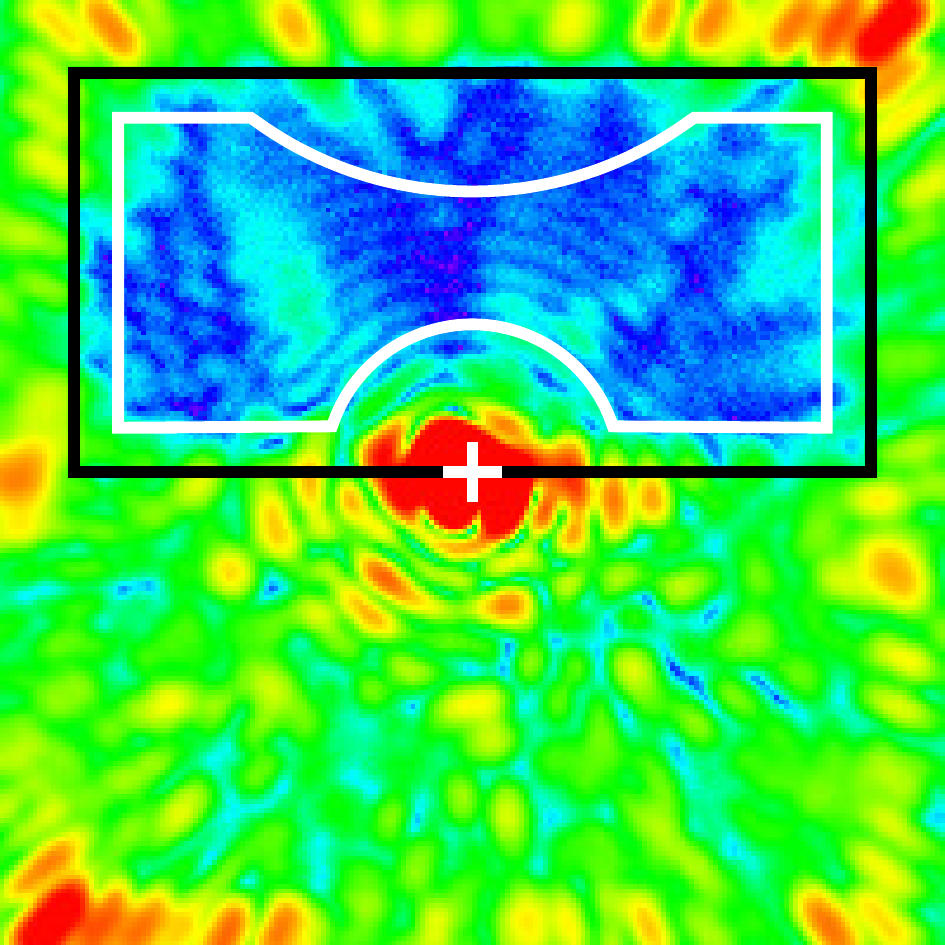}
        \end{subfigure}  
        \caption{Numerical simulated image obtained after correction using a MRSCC in polychromatic light $\Delta\lambda = 80\:\textrm{nm}$ and assuming the optical configuration of the THD bench. Field of view: 32 by 32 $\lambda_{0}/D$. }
        \label{Fig_THD_im_simu}
\end{figure}

Figure \ref{Fig_THD_im_simu} presents a coronagraphic image obtained by using the MRSCC with $\Delta\lambda = 80$ nm. The correction is efficient almost everywhere in the DH.
\begin{figure}
        \centering
        \begin{subfigure}[b]{0.36\textwidth}
        \includegraphics[trim = 11mm 6mm -10mm 8mm, clip, width = 1.1 \textwidth] {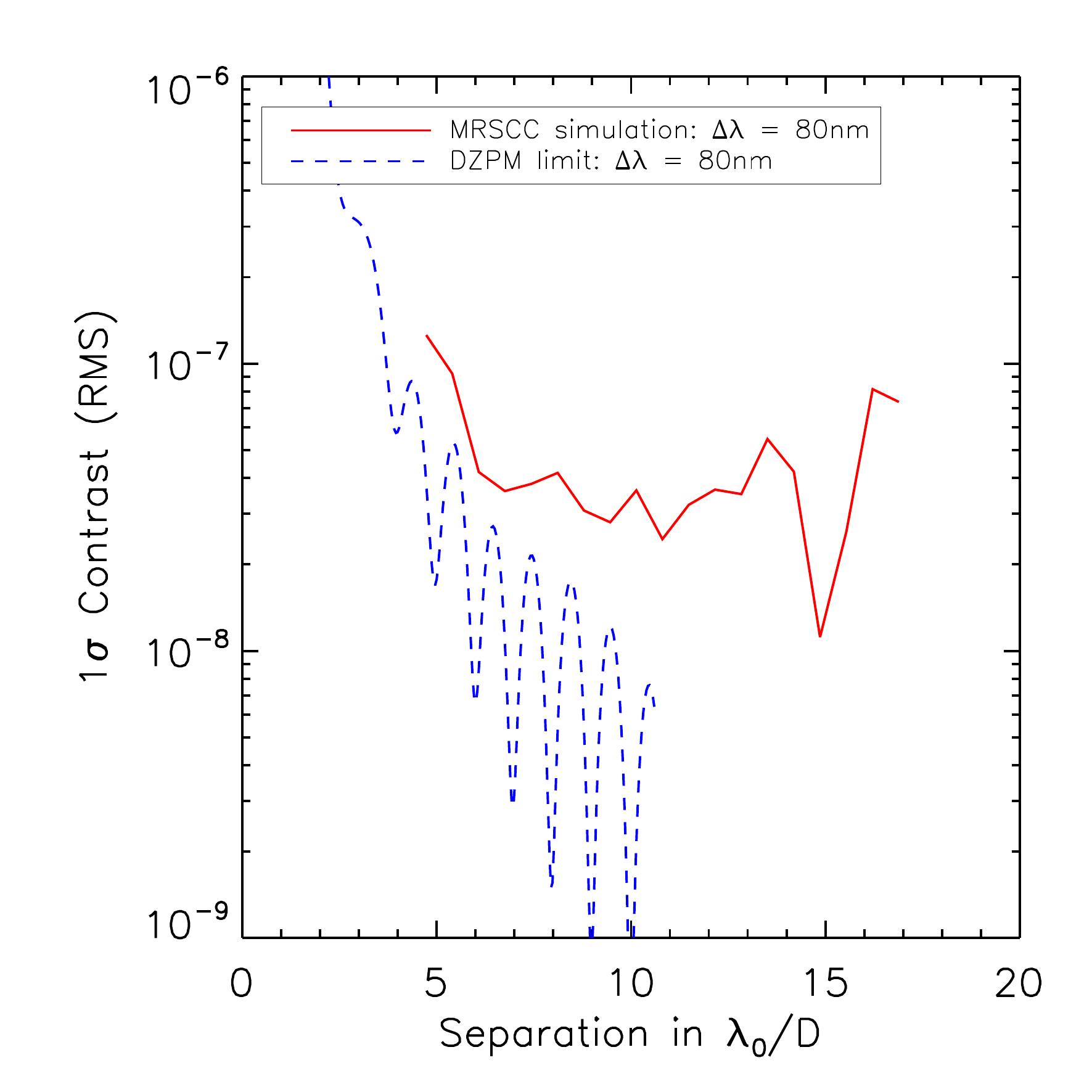}
        \end{subfigure}

        \begin{subfigure}[b]{0.36\textwidth}
        \includegraphics[trim = 11mm 6mm -10mm 8mm, clip, width = 1.1 \textwidth]{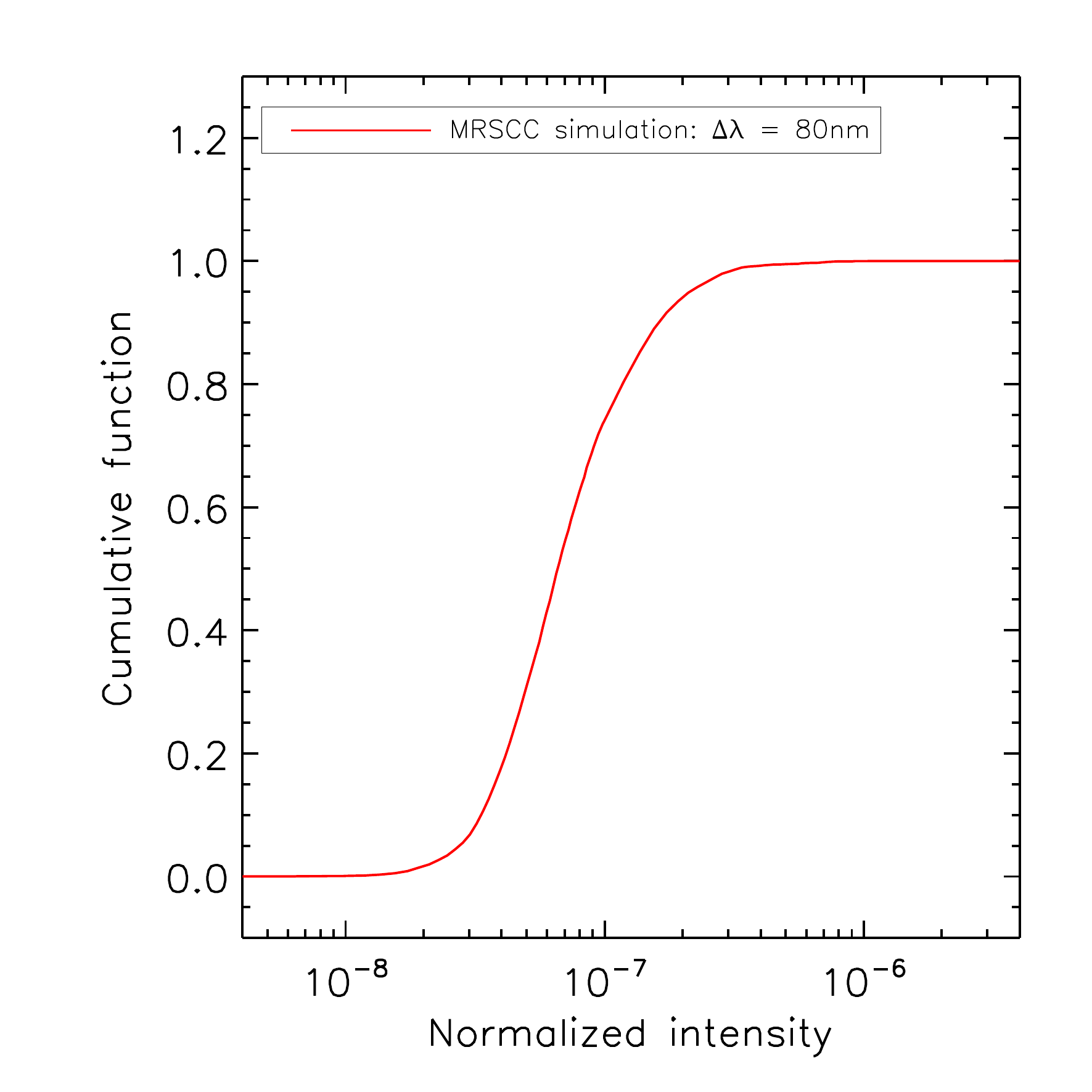}
        \end{subfigure}
        \caption{Contrast curve (top) and spatial cumulative function (bottom) associated with the MRSCC image obtained by numerical simulations assuming the optical configuration of the bench. The blue dashed curve is the theorerical limit of the DZPM coronagraph for a bandwidth of $\Delta\lambda = 80$nm around $\lambda_{0} = 640$nm.}
        \label{THD_im_simu}
\end{figure}
Figure \ref{THD_im_simu} presents the associated contrast curve (top) and the cumulative function (bottom). Below 5$\lambda_{0}/D$ contrast is limited by DZPM coronagraph performance (blue dashed curve). In these numerical simulations, the MRSCC reaches a 1$\sigma$  contrast level of $3.6\:10^{-8}$ between 5$\lambda_{0}/D$ and 17$\lambda_{0}/D$ and a median of the speckle intensity of $6.5\:10^{-8}$. Thus, we expect the MRSCC to significantly reduce the speckle intensity inside the DH in polychromatic light in our laboratory. 

\section{Laboratory performances}
\label{Sec_THD}

\subsection{Chromatism of the bench}
\label{Sec_THD_Sub_Results_Sub_chromatism}
\begin{figure}
        \centering
        \begin{subfigure}[b]{0.36\textwidth}
        \includegraphics[width = \textwidth]{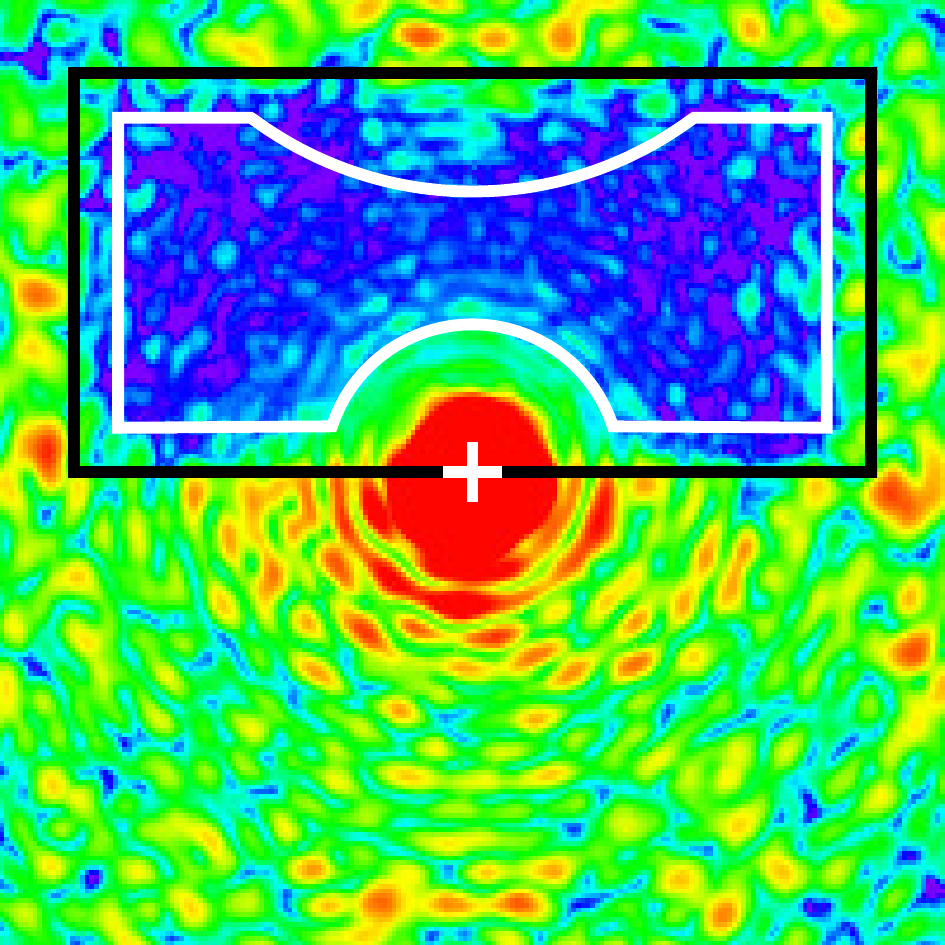}        
        \end{subfigure}
        
        \begin{subfigure}[b]{0.36\textwidth}
        \includegraphics[width = \textwidth]{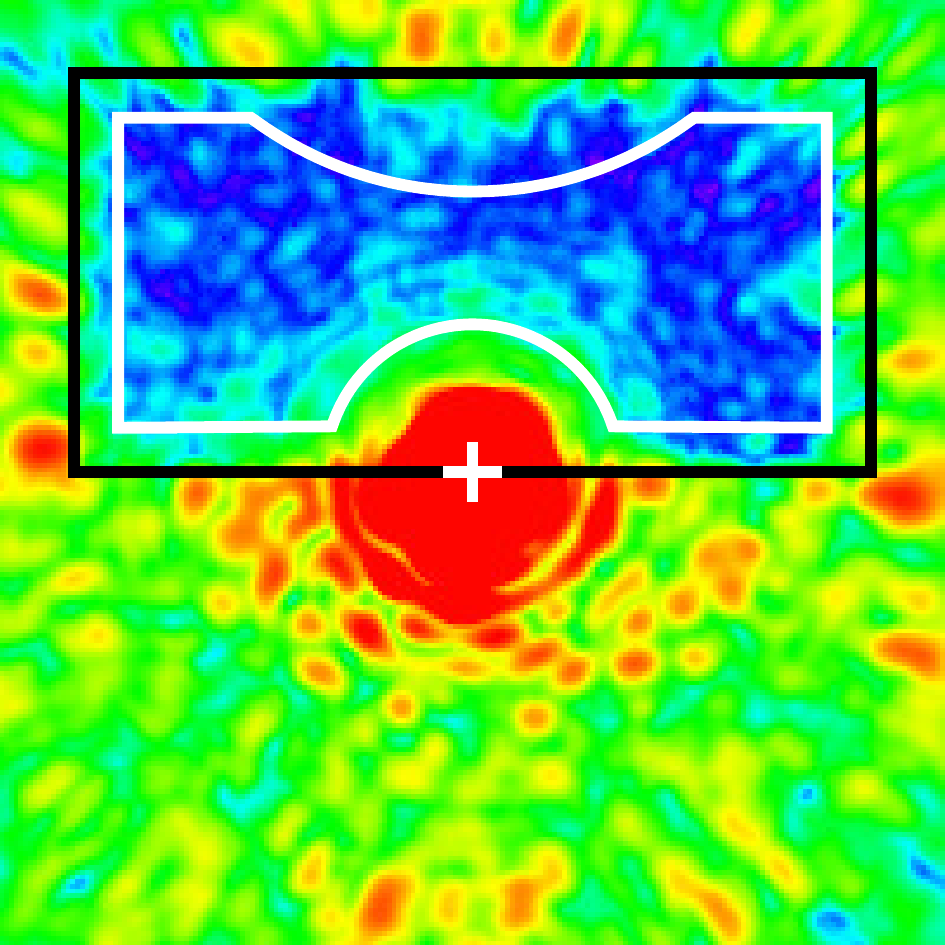}
        \end{subfigure}
        \caption{Laboratory images obtained after SCC correction of the speckles in monochromatic light, (top) recorded in monochromatic light ($\lambda_{0} = 640\:\textrm{nm}$) and  (bottom) recorded in polychromatic light ($\Delta\lambda = 80\:\textrm{nm}$ at $\lambda_{0} = 640\:\textrm{nm}$).}
        \label{Fig_ultima_1}
\end{figure}

\begin{figure}
        \centering
     \begin{subfigure}[b]{0.36\textwidth}
                \includegraphics[trim = 11mm 6mm -10mm 8mm, clip, width = 1.1 \textwidth]{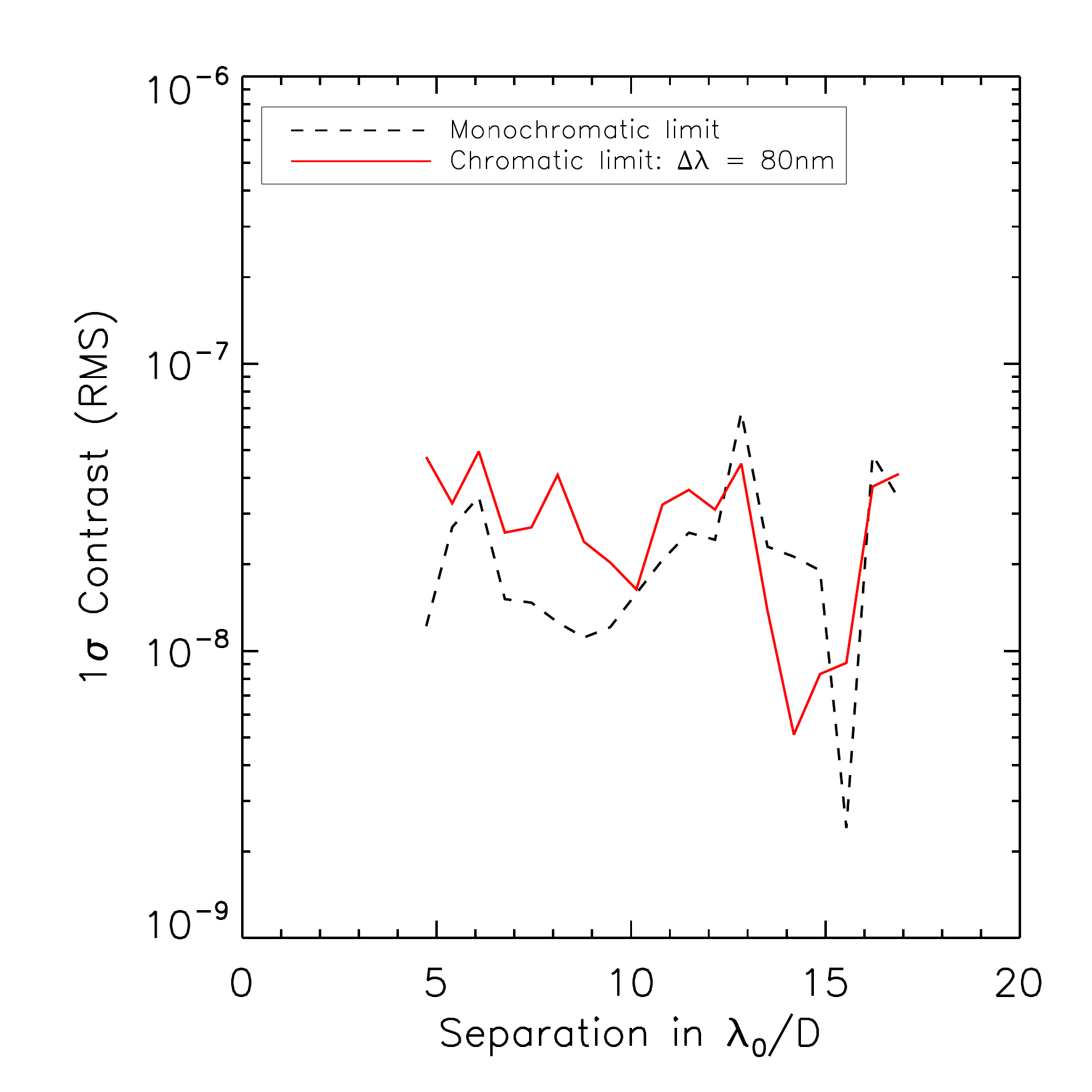}                
        \end{subfigure}
        
        \begin{subfigure}[b]{0.36\textwidth}
        \includegraphics[trim = 11mm 6mm -10mm 8mm, clip, width = 1.1 \textwidth]{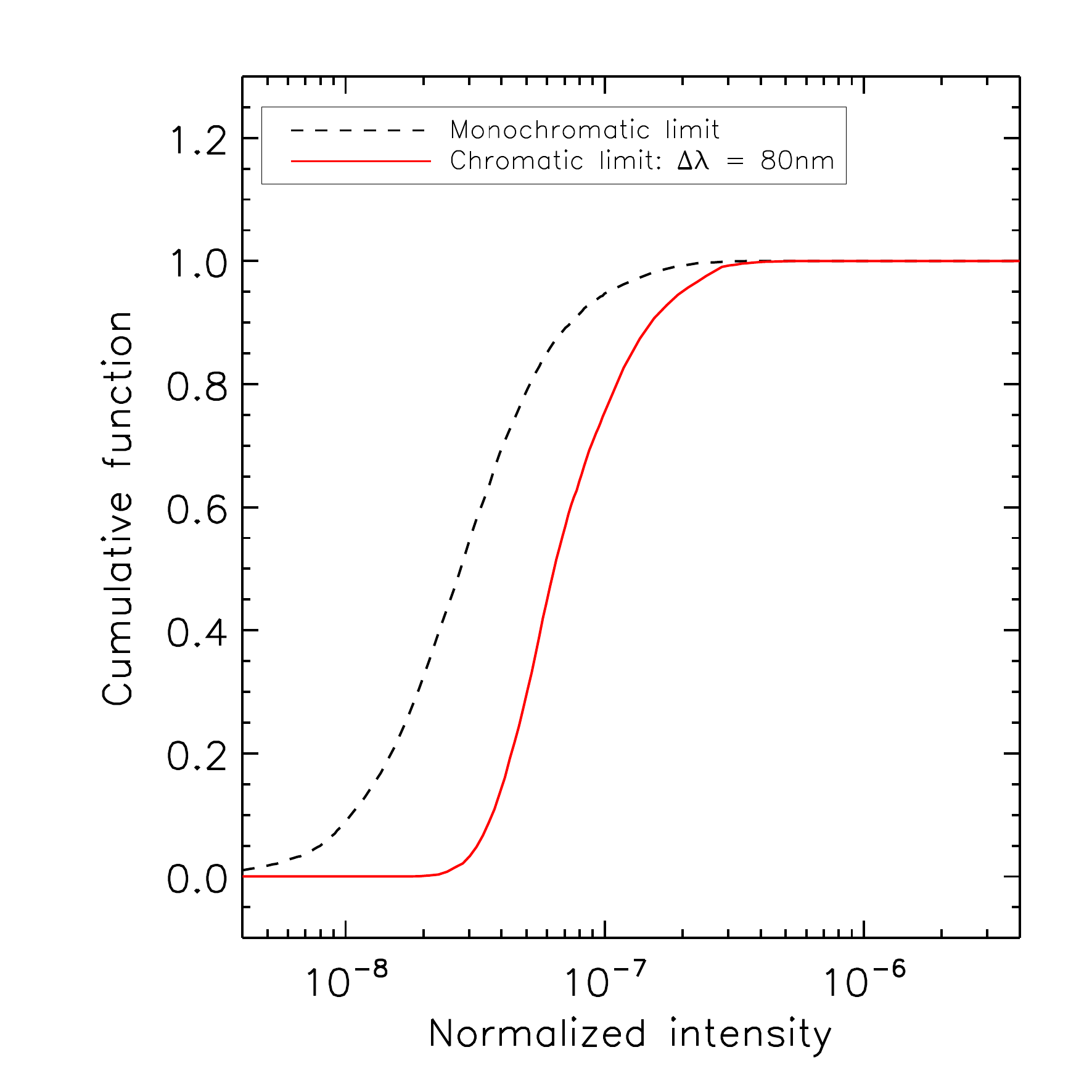}
        \end{subfigure}
        \caption{Contrast curve (top) and spatial cumulative function (bottom) associated with the laboratory images recorded in monochromatic light (black dashed curve) and in polychromatic light (red solid curve)}
        \label{THD_im}
\end{figure}

Before testing the SCC and the MRSCC in polychromatic light, we estimate the chromatism of the optical bench. To obtain this information we use the SCC in monochromatic light (laser source - $\lambda_{0} = 637\: \textrm{nm} - \Delta\lambda<1\:\textrm{nm}$) to reach the deepest contrast that our bench can provide inside the DH. After the correction in closed loop, we record the image, which is presented in Fig. \ref{Fig_ultima_1} (top). Then, we keep the DM shape frozen and we switch the light source from the monochromatic light to the polychromatic light ($\lambda_{0} = 640\:\textrm{nm} - \Delta\lambda=80\:\textrm{nm}$). The resulting image is presented in Fig. \ref{Fig_ultima_1} (bottom). The only difference between the two images  is the bandwidth used to record the two images. The two images are very similar and we study their contrast curves and cumulative functions as given in Fig. \ref{THD_im}.

The contrast curves (top) are not plotted below 5 $\lambda_{0}/D$ because the computation area excludes the central saturated spot (Fig. \ref{Fig_THD_DH}). The average 1$\sigma$ contrast level measured between 5 and 17 $\lambda_{0}/D$ is $2.1\:10^{-8}$ for the image recorded in monochromatic light and $2.7\:10^{-8}$ for the image recorded in polychromatic light.  If we only consider the contrast curves, we conclude that there is no significant chromatic evolution of the aberrations on the THD bench below a contrast of $2.7\:10^{-8}$.

The cumulative functions are plotted at the bottom. The median intensity of the speckles is $2.7\:10^{-8}$ for the image recorded in monochromatic light and $6.3\:10^{-8}$ for the image recorded in polychromatic light. The difference between the two cumulative curves is larger than that between the contrast curves. It means that uniform offset in intensity exists in the polychromatic image. This may be due to chromatism effects of the DZPM, or the outband of the spectral filter or chromatism of the THD optics. We conclude that the THD bench (optical components and coronagraph) is achromatic at a $6.3\:10^{-8}$  level in normalized intensity and at a $2.7\:10^{-8}$ level contrast. These levels give an idea of the deepest contrast that we can achieve in polychromatic light ($\Delta\lambda = 80\:\textrm{nm}$ around $\lambda_{0} =640\:\textrm{nm}$) on the THD bench, whichever technique is used to control the DM.

\subsection{SCC vs MRSCC in polychromatic light}
\label{Sec_THD_Sub_Results_Sub_SCC_vs_MRSCC}

\begin{figure}
        \centering
        \begin{subfigure}[b]{0.36\textwidth}   
        \includegraphics[width = \textwidth]{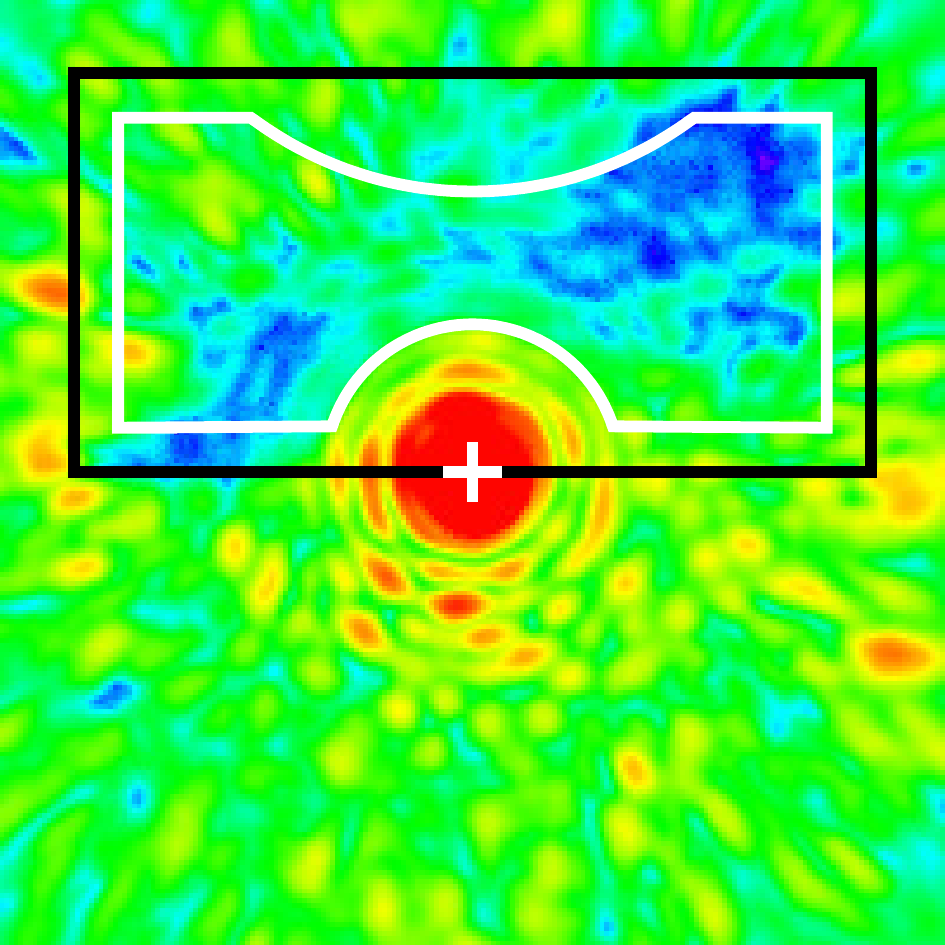}
        \end{subfigure}
        
        \begin{subfigure}[b]{0.36\textwidth}        
        \includegraphics[width = \textwidth]{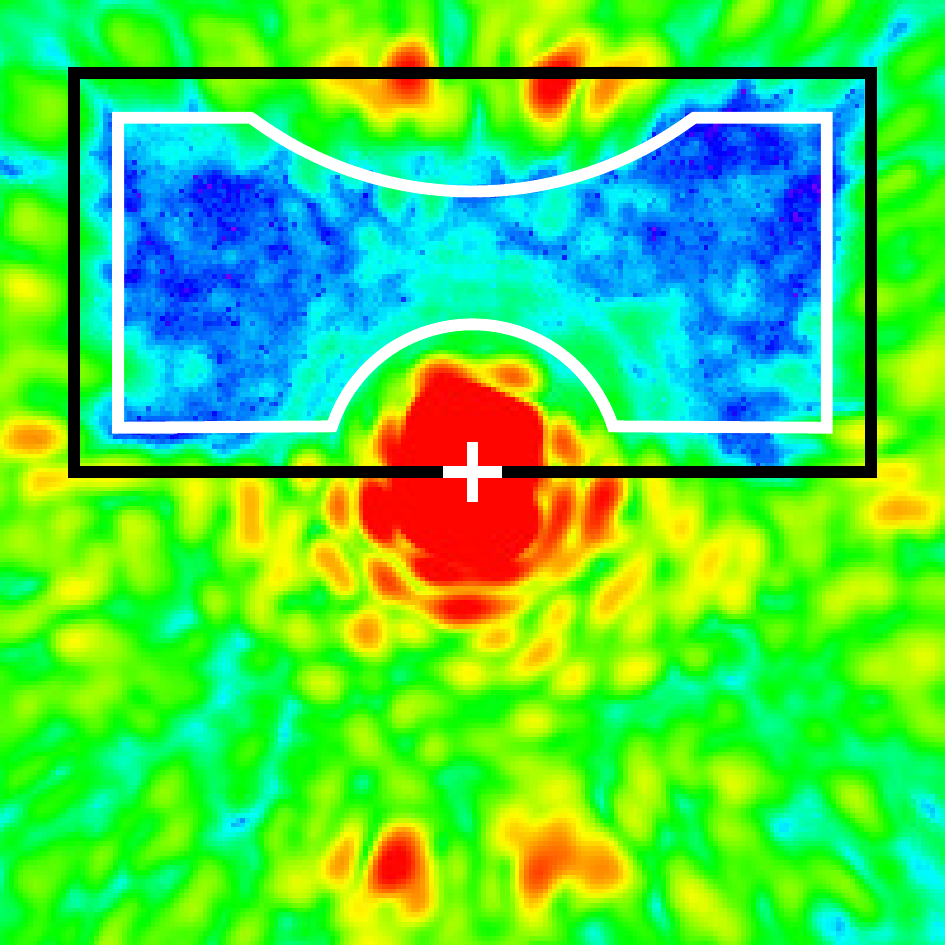}
        \end{subfigure}         
        \caption{Laboratory image obtained on the THD bench after speckle correction in polychromatic light ($\Delta\lambda = 80$ nm at $\lambda_{0} = 640$ nm) using a SCC (top) and the MRSCC (bottom).} 
        \label{Fig_SCC_THD}
\end{figure}

\begin{figure}
        \centering
        \begin{subfigure}[b]{0.36\textwidth}
        \includegraphics[trim = 11mm 6mm -10mm 8mm, clip, width = 1.1 \textwidth] {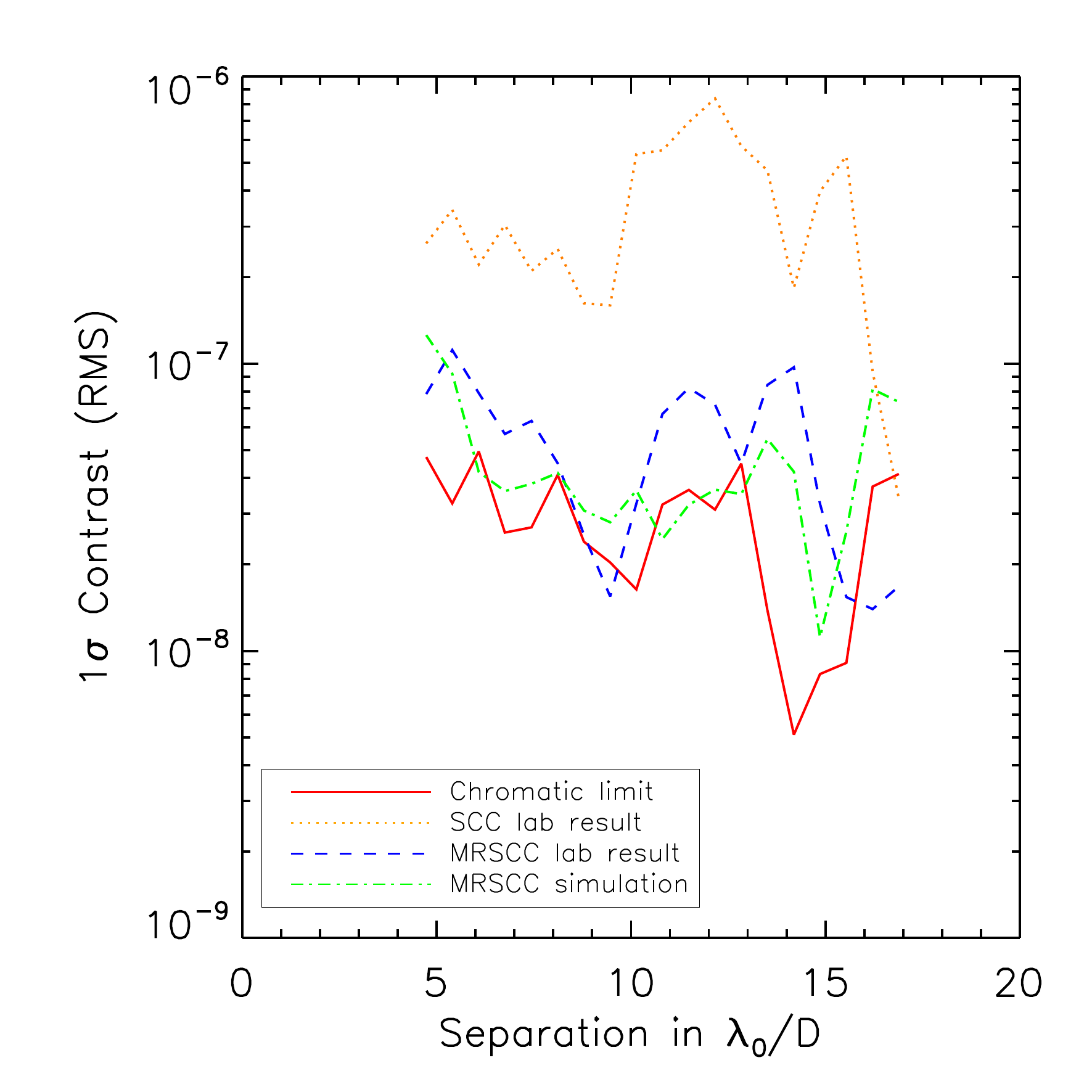}
        \end{subfigure}
        \begin{subfigure}[b]{0.36\textwidth}
        \includegraphics[trim = 11mm 6mm -10mm 8mm, clip, width = 1.1 \textwidth]{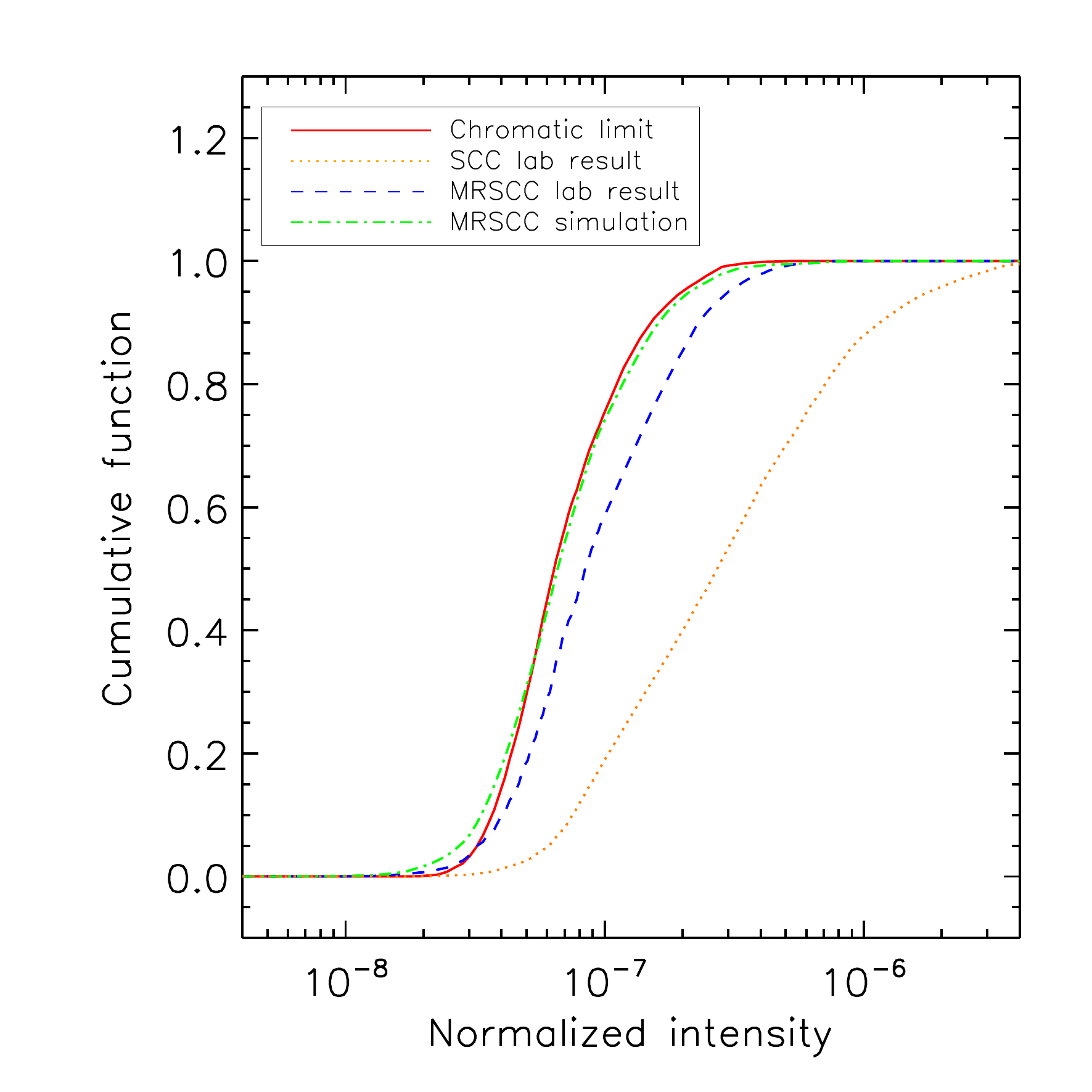}
        \end{subfigure}
        \caption{Contrast curves (top) and spatial cumulative functions (bottom) for $\Delta\lambda = 80$ nm, associated with the expected performance from numerical simulations (mixed green line) and to three laboratory images: chromatic limit of the THD bench (red solid line), SCC correction (orange dotted curves) and MRSCC correction (blue dashed curves).The cumulative curves have been computed inside the computation area presented Fig. \ref{Fig_THD_DH}.}
        \label{THD_REDU}
\end{figure}

In this section we compare the SCC and MRSCC performance in the laboratory using a bandwidth of 80 nm around $\lambda_{0} =  640\:\textrm{nm}$. First, we create a DH using the SCC in polychromatic light starting from a given voltage map for the DM (i.e., a given surface shape) and we record the best correction of the speckle field (Fig. \ref{Fig_SCC_THD}, top). Then, starting from the same given initial DM shape, we use the MRSCC and we obtain the DH of Fig. \ref{Fig_SCC_THD} (bottom). As expected, in the SCC image, a lot of speckles are not corrected, whereas the correction is almost uniform with the MRSCC.
Figure \ref{THD_REDU} presents the contrast (top) and cumulative (bottom) curves computed from these images. The orange dotted curves are associated with the laboratory image obtained with the SCC (Fig. \ref{Fig_SCC_THD}, top) and the blue dashed curves to the laboratory image obtained with the MRSCC (Fig. \ref{Fig_SCC_THD}, bottom). We overplot the curves associated with the chromatic limit of the bench (red solid curves, Sect. \ref{Sec_THD_Sub_Results_Sub_chromatism}) and associated with the numerical simulations of the bench (mixed green curves, Section \ref{SubSec_TR}).

All these laboratory results demonstrate the gain in contrast brought by the MRSCC compared to the SCC. For SCC, the average 1$\sigma$  contrast is never better than $10^{-7}$ and the median intensity is $2.7\:10^{-7}$. In the MRSCC images we measure an average 1$\sigma$  contrast of $4.5\:10^{-8}$ between 5 and 17 $\lambda_{0}/D$ and a median intensity of $8.3\:10^{-8}$. These performances are close to the numerical simulations prediction: $3.6\:10^{-8}$ and $6.5\:10^{-8}$ respectively.

If we compare the images obtained with the MRSCC (Fig. \ref{Fig_SCC_THD}, bottom) and the image that is limited by the chromatism of the bench (Fig. \ref{Fig_ultima_1}, bottom), the levels are equivalent, except at the top of the DH, where some speckles are not well corrected because all the fringe patterns of the MRSCC are blurred. These types of  bright uncorrected speckles also exist in the numerical simulated image (top right corner of the DH in Fig. \ref{Fig_THD_im_simu}). From numerical simulations we find that the location of these speckles depends on the initial phase and amplitude aberrations. It may be possible to correct them by adding a fourth reference hole or by optimizing the orientations of the reference holes. However, none of these solutions were available during our laboratory experiments 

\section{Conclusions}
\label{Conclusions}

In this paper, we proposed two methods to make the SCC more achromatic: the MRSCC and the OPD method. We tested these two methods by numerical simulations and proved that they improve the performance of the SCC in polychromatic light. We found that the OPD method performance depends on the initial aberration map and that it is not robust enough to be used.

Then, we demonstrated in the laboratory that the MRSCC can be used to control a DM in polychromatic light (bandwidth of 80 nm around 640 nm - 12.5\%) to reach an average 1$\sigma$  contrast of approximately $4.5\:10^{-8}$  between 5 and 17 $\lambda_{0}/D$ from the star  and a median contrast of $8.3\: 10^{-8}$. The performances obtained on the THD bench with the DZPM coronagraph are close to the chromatic limitation of our bench, meaning that the MRSCC may be able to control the speckle intensity at even deeper contrast levels.

The MRSCC has all the advantages of the SCC. It is a focal plane wavefront sensor, which is able to work in closed loop. It is easy to implement by adding only small holes in the Lyot stop plane and it can be associated with several phase-mask coronagraphs. Also, only one focal plane image is mandatory for estimating the electric field instead of the several images that are needed by techniques that are based on temporal modulation. 

This study was realized using a clear circular aperture. In the case of an arbitrary telescope aperture, the sensing with the MRSCC can still be undertaken as long as the coronagraph diffracts enough stellar light into the reference hole.

The attractive features of the MRSCC make this FPWFS a serious candidate for a possible upgrade of the current high-contrast imaging instruments and could be used with future instruments for space-based missions and extremely large telescopes.

Finally, several improvements may be possible in the future. Among them, we will explore more sophisticated algorithms to measure the electric field  from the MRSCC image. We also indend to correct speckles close to the optical axis by using another phase-mask coronagraph like a vortex \citep{Mawet2005}. Finally, \cite{Mazoyer2013a} demonstrate that the contrast level of the THD bench is determined mainly by amplitude aberrations. To improve its performance and increase the size of the DH, we will control both the  phase and amplitude aberrations in the full influence area with an upgraded version of our bench by adding DMs. 
 
\begin{acknowledgements}
We wish to thank M. N'Diaye, A. Caillat and K. Dohlen for their collaboration on the design of the DZPM used to obtain laboratory results and for sharing the code used to numerically simulate it. This work was carried out at the Observatoire de Paris (France) under contract number DA-10091195 with the CNES (Toulouse, France). We also thank the referee for his very constructive remarks on the manuscript of the paper.
\end{acknowledgements}

\bibliographystyle{aa.bst}
\bibliography{report.bib}   

\end{document}